\documentclass{article}
\usepackage{graphicx} 
\usepackage[top=2cm, bottom=2cm, left=2cm, right=2cm]{geometry}
\usepackage[caption=false]{subfig}
\usepackage{epsfig}
\usepackage[toc,page]{appendix}
\usepackage{amsmath}
\usepackage{mathtools}

\begin{document}
\begin{center}
{\huge \bfseries Ion temperature gradient mode mitigation by energetic particles, mediated by forced-driven zonal flows \\[0.4cm]}
J.N Sama$^1$, A. Biancalani$^2$, A. Bottino$^3$, D. Del Sarto$^1$,R. J. Dumont$^4$, G. Di Giannatale$^5$, A. Ghizzo$^1$, T. Hayward-Schneider$^3$, Ph. Lauber$^3$, B. McMillan$^6$, A. Mishchenko$^7$, M. Muruggapan$^5$, B. Rettino$^3$, B. Rofman$^5$, F. Vannini$^3$, L. Villard$^5$  and X. Wang$^3$\\
$^1$Université de Lorraine, CNRS, IJL, Nancy, France,\\
$^2$Léonard de Vinci Pôle Universitaire, Research Center, Paris La Défense, France,\\
$^3$Max Planck Institute for Plasma Physics, Garching, Germany,\\
$^4$CEA, IRFM, F-13108 Saint-Paul-lez-Durance, France,\\
$^5$Swiss Plasma Center, EPFL, Lausanne, Switzerland,\\
$^6$Center for Fusion, Space and Astrophysics, University of Warwick, Coventry, UK,\\
$^7$Max Planck Institute for Plasma Physics, Greifswald, Germany,\\

\end{center}
\section*{Abstract}
In this work, we  use the global electromagnetic and electrostatic gyro kinetic approaches to investigate the effects of zonal flows forced-driven by Alfvén modes due to their excitation by energetic particles (EPs), on the dynamics of ITG (Ion temperature gradient) instabilities. The equilibrium of the 92416 JET tokamak shot is considered. The linear and nonlinear Alfvén modes dynamics, as well as the zonal flow dynamics, are investigated and their respective radial structures and saturation levels are reported. ITG dynamics in the presence of the zonal flows excited by these Alfvén modes are also investigated. We find that, the zonal flows forced-driven by Alfvén modes can significantly impact the ITG dynamics. A zonal flow amplitude scan reveals the existence of an inverse relation between the  zonal flow amplitude and the ITG growth rate. These results show that, forced-driven zonal flows can be an important indirect part of turbulence mitigation due to the injection of energetic particles. 
\section{Introduction}
Drift wave turbulence and its associated heat and particle transport have been intensively studied due to the role they play in deteriorating plasma confinement, which is a concern for future fusion plasma reactors. The ion-temperature gradient modes (ITG) are drift waves driven unstable by the ion-temperature gradient. The spontaneous non-linear excitation of zonal flows by ITG modes \cite{hasegawa,rosenblut,zonca}, is believed to be one of their major saturation mechanisms. Zonal flows (ZF) i.e. axisymmetric perturbations of tokamak plasmas, play an important role in the self-consistent saturation of turbulence. There are two types: the zero frequency zonal flow (ZFZF) and the finite frequency geodesic acoustic modes (GAMs)\cite{winsor,ales_gam,Ivan,sama}. The ZFZF can be generated both by ITGs and by their lower frequency branch, in which only ions trapped in the banana orbits intervene\cite{ghizzo,Aghizzo,Drouot,Gravier,Ghizzo_2017}. Trapped Ion Modes (TIM)  is the name given to this lower frequency branch of ITGs (an electron counterpart also exists). The  turbulence and ZFZF generated by the high frequency branch of ITGs are generally dominant over those generated by the TIMs. The ZFZF, driven by drift waves like ITGs via modulational instability, is dominated by the Reynolds and Maxwell stresses due to thermal species non-linearity \cite{chen2000}.

It has been observed experimentally in some tokamaks such as the JET \cite{Romanelli} and AUG \cite{Tardini} that under certain conditions, the injection of a small population of energetic particles, can lead to the transition of the tokamak plasma, from a low confinement regime, characterized by high transport and turbulence, known as the L-mode to a high confinement regime characterized by a significant reduction of turbulence and transport, known as the H-mode \cite{wagner}. Zonal flows are believed to play a key role in this transition, however, the exact source of these zonal flows and the role played by energetic particles still remains an open question. Several analytical theories and models \cite{white,Zonca_2015,ningfei,Hahm,Ghizzo_2023} have been developed in the last decades and numerical flux-tube simulations performed \cite{Angioni_2009,em,zhang,citrin,Garcia_2015,Di}. Qiu and co-workers \cite{qiu}, proposed a theoretical model in which zonal flows can be driven by Alfvén modes (AM)\cite{Appert_1982,zonca2} in the presence of energetic particles. According to their model, the energetic particles non-linearity that enters the vorticity equation through the curvature coupling term, dominates over the bulk particles non-linearity associated to the Reynolds and Maxwell stresses responsible for the excitation of zonal flows by drift waves. Under this condition, the Alfvén modes can drive a zonal flow with a growth rate that is twice that of the Alfvén modes. The zonal flows excited by this mechanism are said to be forced-driven.
 
\begin{figure}[ht]
\centering
\includegraphics[trim=5cm 0 5cm 0,clip,width=6.5cm]{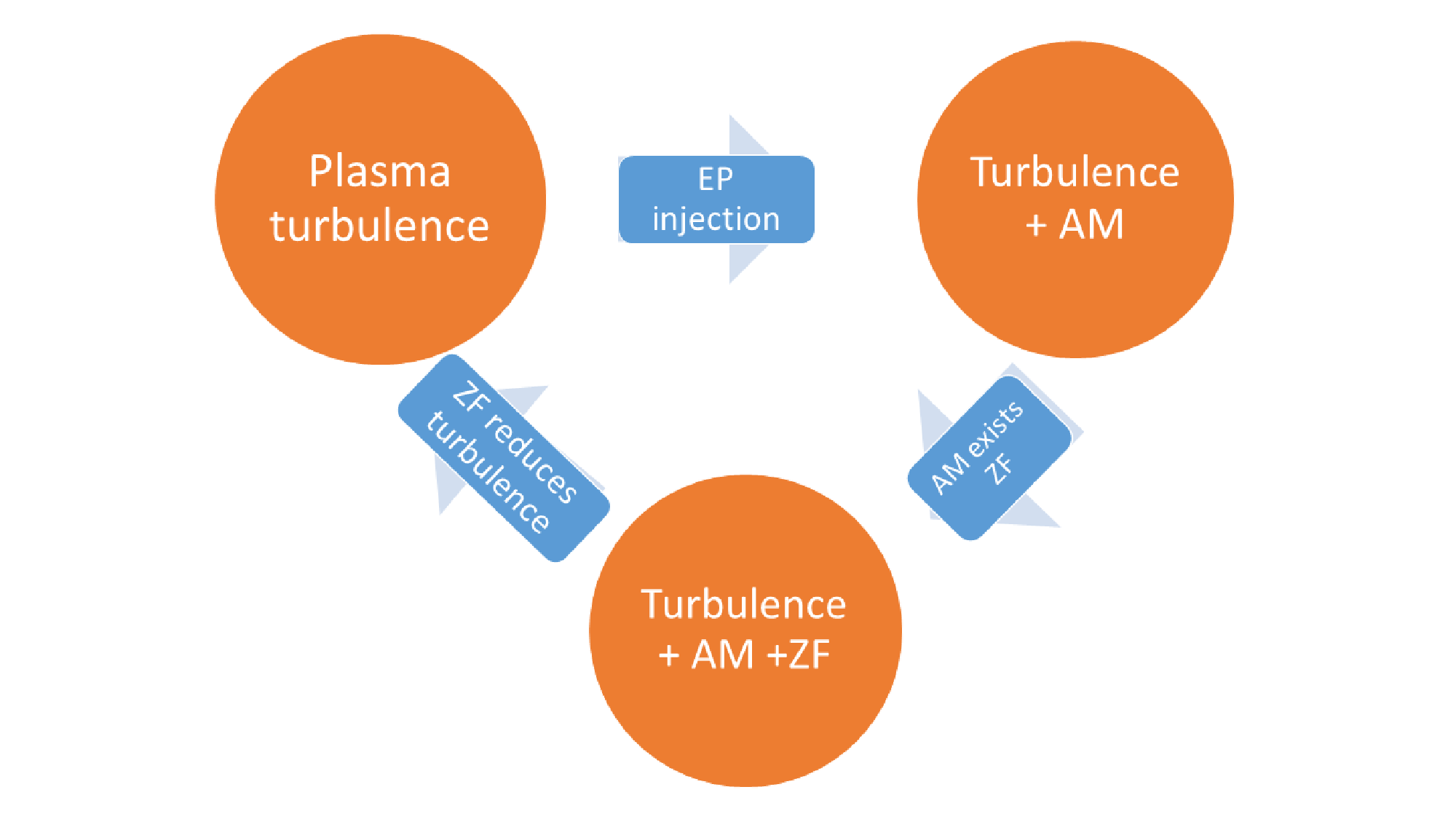}
\caption{Cartoon summarizing the conjecture we aim to verify with the numerical simulations  discussed herein.}
\label{fig1}
\end{figure}

 Recently, global simulations with the gyro-kinetic particle-in-cell code ORB5 \cite{MISHCHENKO2019194,Lanti} have been performed, where the self-consistent interaction of ITG turbulence, zonal flows, and Alfvén modes has been studied in the presence of energetic particles \cite{biancalani_bottino_lauber_mishchenko_vannini_2020,alessandro,biancalani_bottino_del}. 
 Similar studies have been made with other codes \cite{siena,Ishizawa_2021}. As a result, it has been conjectured that zonal flows force-driven by Alfvén instabilities may be used to reduce the level of ITG turbulence.This conjecture proposes a possible mechanism, where the indirect interaction between energetic particles and turbulence is mediated by forced-driven zonal flows, as illustrated in Fig. \ref{fig1}. 

In this work, we design numerical experiments, to test in experimentally relevant scenarios part of the above conjecture that claims that the  forced-driven zonal flows can significantly mitigate ITG instability. To achieve this objective, we isolate this mechanism from self-consistent nonlinear simulations, and we prove that it is indeed effective by measuring the amplitude and radial structure of the force-driven zonal flows, and by imposing this radial structure on electrostatic ITG simulations.  Imposing this self-consistent zonal flows leads to a significant mitigation of the ITG instability, corresponding to a drop of its growth rate. A zonal flow amplitude scan reveals the existence of inverse relation between the zonal flow amplitude and the ITG growth rate.

This work is divided into four sections: An introductory section in which the motivations of the work are given. In section two, we describe the ORB5 numerical scheme, while  the dynamics of Alfvén modes and their nonlinear excitation of zonal flows is presented in section three. The dynamics of ITG and forced-driven zonal flows is studied in section four. 
\section{Model}
 The main numerical tool used for the investigations described herein is the global electromagnetic particle-in-cell code ORB5. The model equations of ORB5 derived using the mixed-variable gyro kinetic formalism  are the gyro center trajectories, equation (\ref{1}) and (\ref{2}), quasi-neutrality equation, equation (\ref{3}), Ampère's equations, equation (\ref{4}) and the Ohm's law, equation \ref{5}.
 \begin{equation}
     \frac{d\textbf{R}}{dt}=v_{||}\textbf{b}_0^*+\frac{1}{eB_{||}^*}\textbf{b}\times \mu\nabla B+\frac{\textbf{b}}{B_{||}^*}\times \nabla\left<\phi -v_{||}A_{||}^{(s)}-v_{||}A_{||}^{(h)}\right>-\frac{e}{m}\left<A_{||}^{(h)}\right>\textbf{b}_0^*
     \label{1}
 \end{equation}
 \begin{equation}
     \frac{d\textbf{v}_{||}}{dt}=-\frac{\mu}{m}\textbf{b}_0^*\cdot\nabla B -\frac{e}{m}\left[\textbf{b}^*\cdot\nabla\left<\phi -v_{||}A_{||}^{(h)}\right>+\frac{\partial}{\partial t}\left<A_{||}^{(s)}\right>\right]-\frac{\mu}{m}\frac{\textbf{b}\times\nabla B}{B_{||}^*}\cdot \nabla\left< A_{||}^{(s)}\right>
     \label{2}
 \end{equation}
 \begin{equation}
     \textbf{b}^*=\textbf{b}_0^*+\frac{\nabla \left<A_{||}^{(s)}\right>\times\textbf{b}}{B_{||}^*}, \quad
     \textbf{b}_0^*=\textbf{b}+\frac{mv_{||}}{eB_{||}^*}\nabla\times \textbf{b}, \quad
     B_{||}^*=B+\frac{mv_{||}}{e}\textbf{b}\cdot\nabla\times \textbf{b}     
 \end{equation}
 \begin{equation}
       -\textbf{$\nabla$}\cdot\left(\frac{n_0 }{B\Omega_i}\textbf{$\nabla$}_{\perp}\phi\right)= \Bar{n}_i-\Bar{n}_e
       \label{3}
 \end{equation}
 \begin{equation}
     \sum_{s=i,e}\frac{\beta_s}{\rho^2_s}A_{||}^{(h)}+\nabla^2_{\perp} A_{||}^{h}=\mu_0\sum_{s=i,e}\Bar{j}_s + \nabla^2_{\perp} A_{||}^{s}
     \label{4}
 \end{equation}
 \begin{equation}
  \frac{\partial}{\partial t}A_{||}^{(s)} +\textbf{b}\cdot\nabla\phi =0  
  \label{5}
 \end{equation}
 The phase-space coordinates are $\textbf{Z} = (\textbf{R}, v_{||},\mu)$, i.e. respectively, the gyro center position, the parallel velocity and the  magnetic moment. The time-dependent fields are the scalar potential, $\phi$ and the parallel component of the vector potential, which is split into a symplectic and Hamiltonian part, $A_{||}=A_{||}^{(s)}+A_{||}^{(h)}$. \textbf{B} and \textbf{b} are the equilibrium magnetic field and magnetic unit vector. $\Bar{n}_s$ and $\Bar{J}_s$ are, respectively, the perturbed density and current, while, $\Omega_i$ is the ion cyclotron frequency. The other quantities are written in standard notation. The gyro average operator is defined by:
 \begin{equation}
     \left<\phi\right>=\frac{1}{2\pi}\int_0^{2\pi}\phi(\textbf{$R+\rho_L$})d\alpha
 \end{equation}
where $\alpha$ here is the gyro angle and $\rho_L = \rho_L(\alpha,\mu)$ is the Larmor radius. The gyro average operator reduces to the zeroth Bessel function $J_0(k_{\perp}\rho_{L,i})$ if we transform it in Fourier space. The gyro average is calculated for all ion species. We take into account finite Larmor radius of ions, and we neglect them for electrons. Hereafter we will label with "$s$" the radial coordinate, which is defined in terms of poloidal flux, $\psi_p$ as $s=\sqrt{\psi_p/\psi_{p,edge}}$. A nonlinear pullback scheme is used in all electromagnetic simulations. This scheme includes the nonlinear flutter terms, whose contribution becomes non-negligible for strongly unstable MHD-type modes\cite{Mishchenko_2021}.
\section{Alfvén modes dynamics in JET 92416}
 The first challenge in this section is to be able to run a global electromagnetic simulation with experimentally measured magnetic equilibrium, species temperature and density profiles, in which Alfvén modes excited by energetic particles drive zonal flows. A good experimental case which we will consider is the JET shot $92416$ afterglow experiment\cite{Dumont_2018,Fitzgerald_2022}. In this shot, $n=4,5,6$ TAEs (Toroidal Alfvén eigen modes) were measured at the time $6.2s$ of the discharge. We shall consider the magnetic equilibrium and profiles at this particular time. For this study, we focus only on the dynamics of the $n=5$ mode. A successful benchmark has been performed for this case without fast particles using the MISHKA  code\cite{Mishka} and ORB5.
\subsection{JET 92416 equilibrium}\label{equi}
\begin{figure}[ht]
 \centering
\subfloat[\centering]{{\includegraphics[trim=8cm 0 8 0,clip,width=8cm]{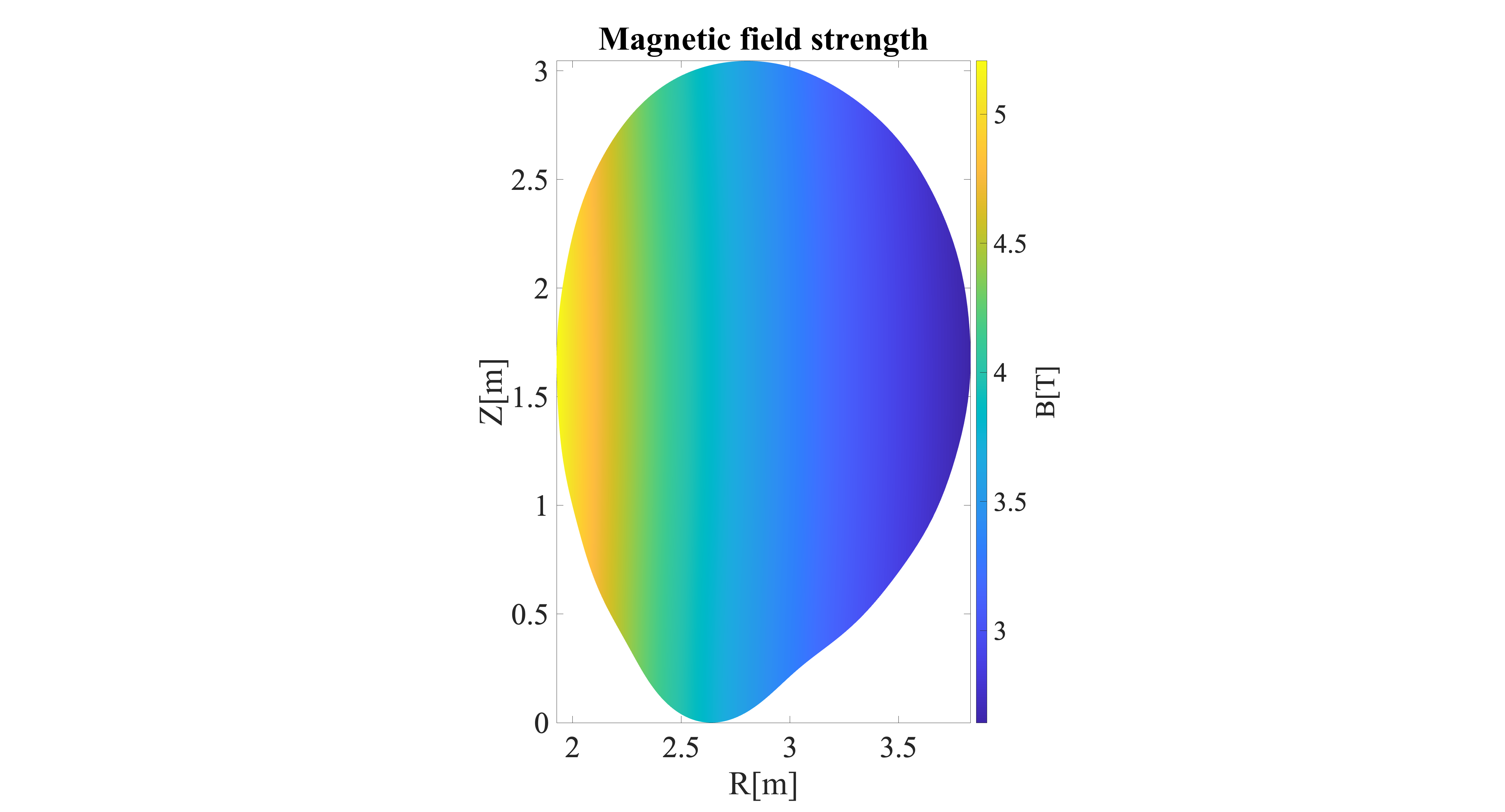}}}
\subfloat[\centering]{{\includegraphics[trim=8cm 0 8 0,clip,width=8cm]{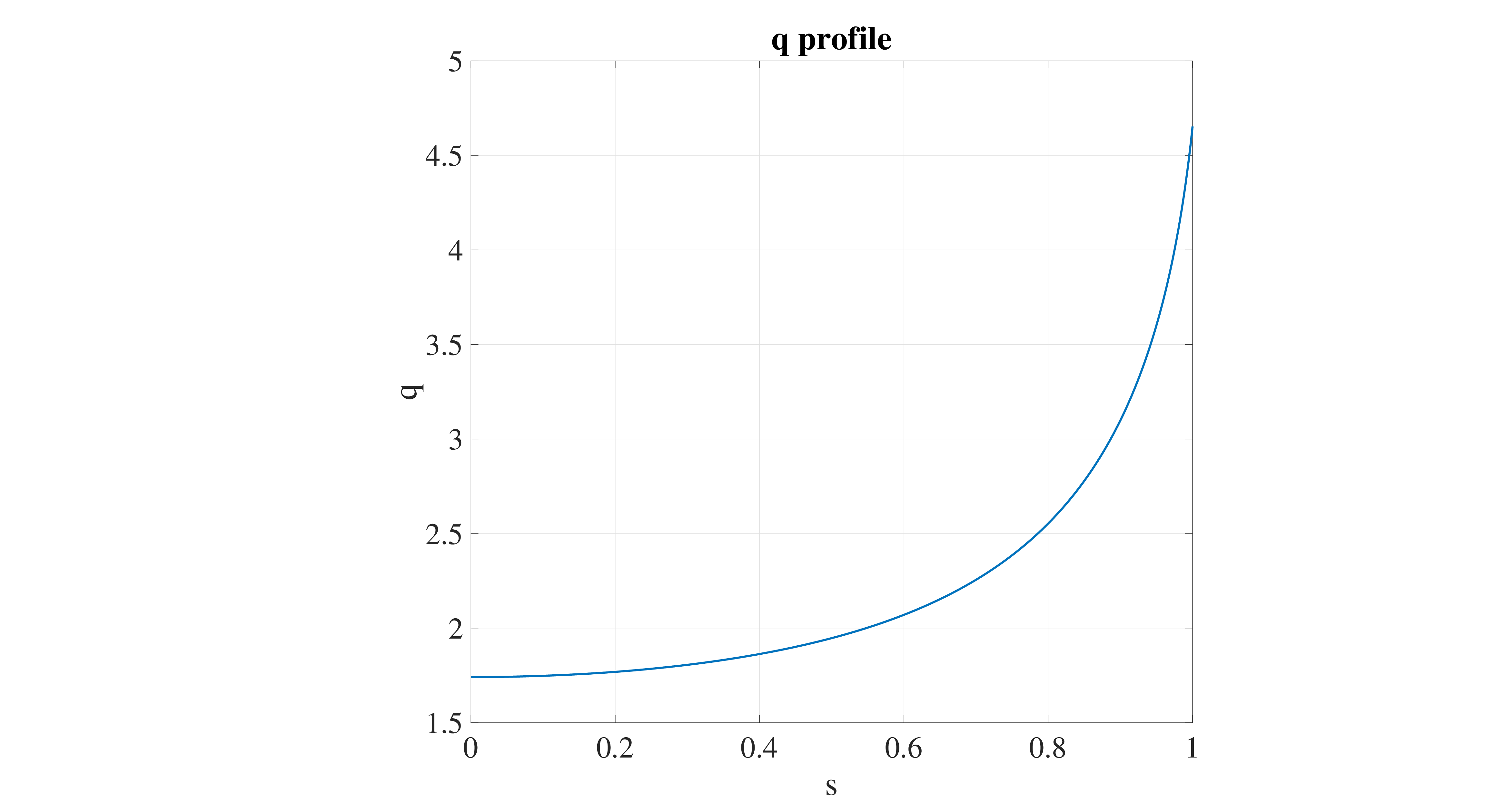}}}
     \caption{Magnetic equilibrium and radial profile deduced from experimental data using the CHEASE code (a) Magnetic field strength (b) Safety factor profile}
     \label{fig9}
 \end{figure}

\begin{figure}[ht]
\subfloat[\centering]{{\includegraphics[trim=8cm 0 8cm 0,clip,width=8.0cm]{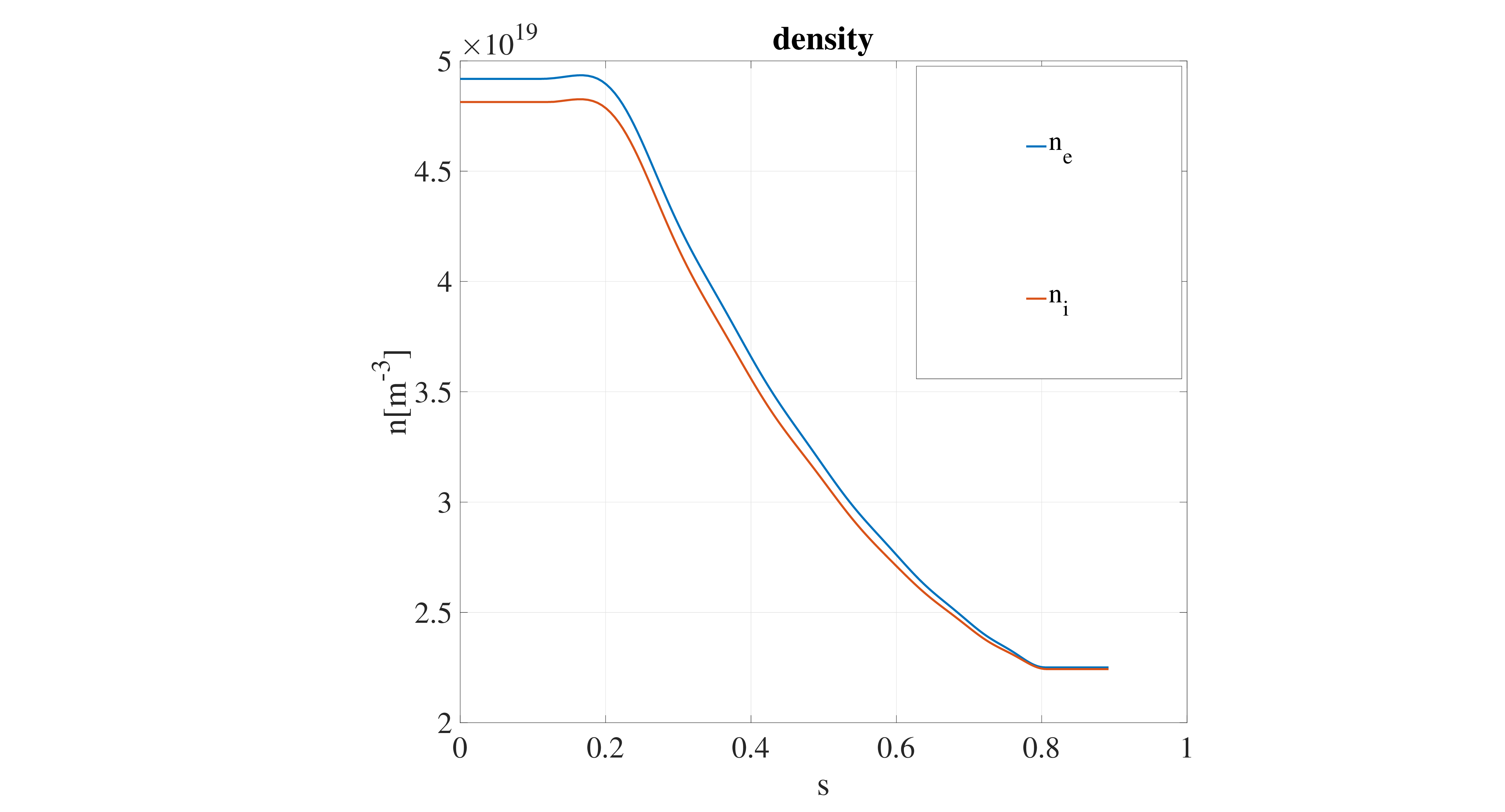}}}
\subfloat[\centering]{{\includegraphics[trim=8cm 0 8cm 0,clip,width=8.0cm]{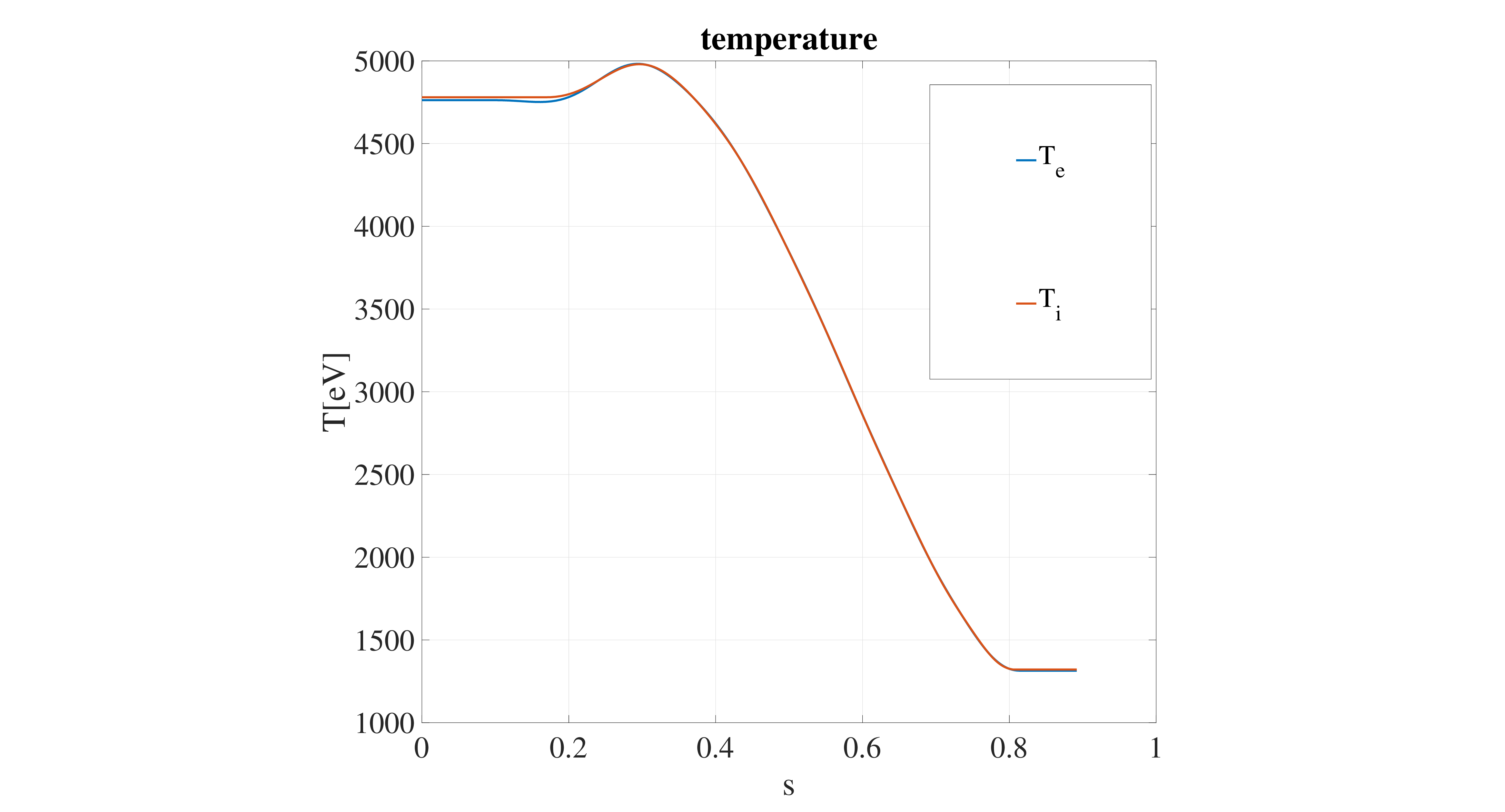}}}
\newline
\subfloat[\centering]{{\includegraphics[trim=8cm 0 12cm 0,clip,width=8cm]{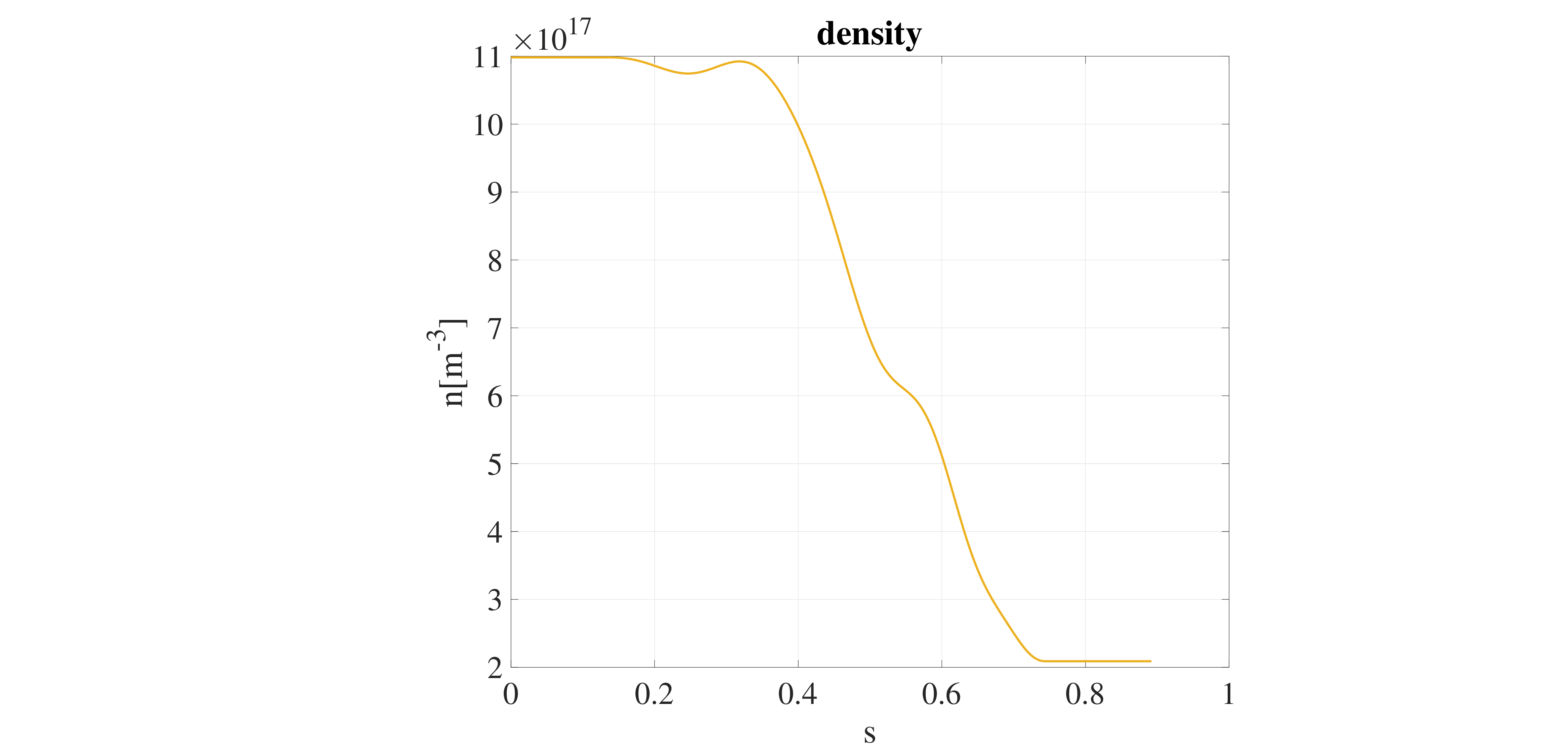}}}
\subfloat[\centering]{{\includegraphics[trim=8cm 0 12cm 0,clip,width=7cm]{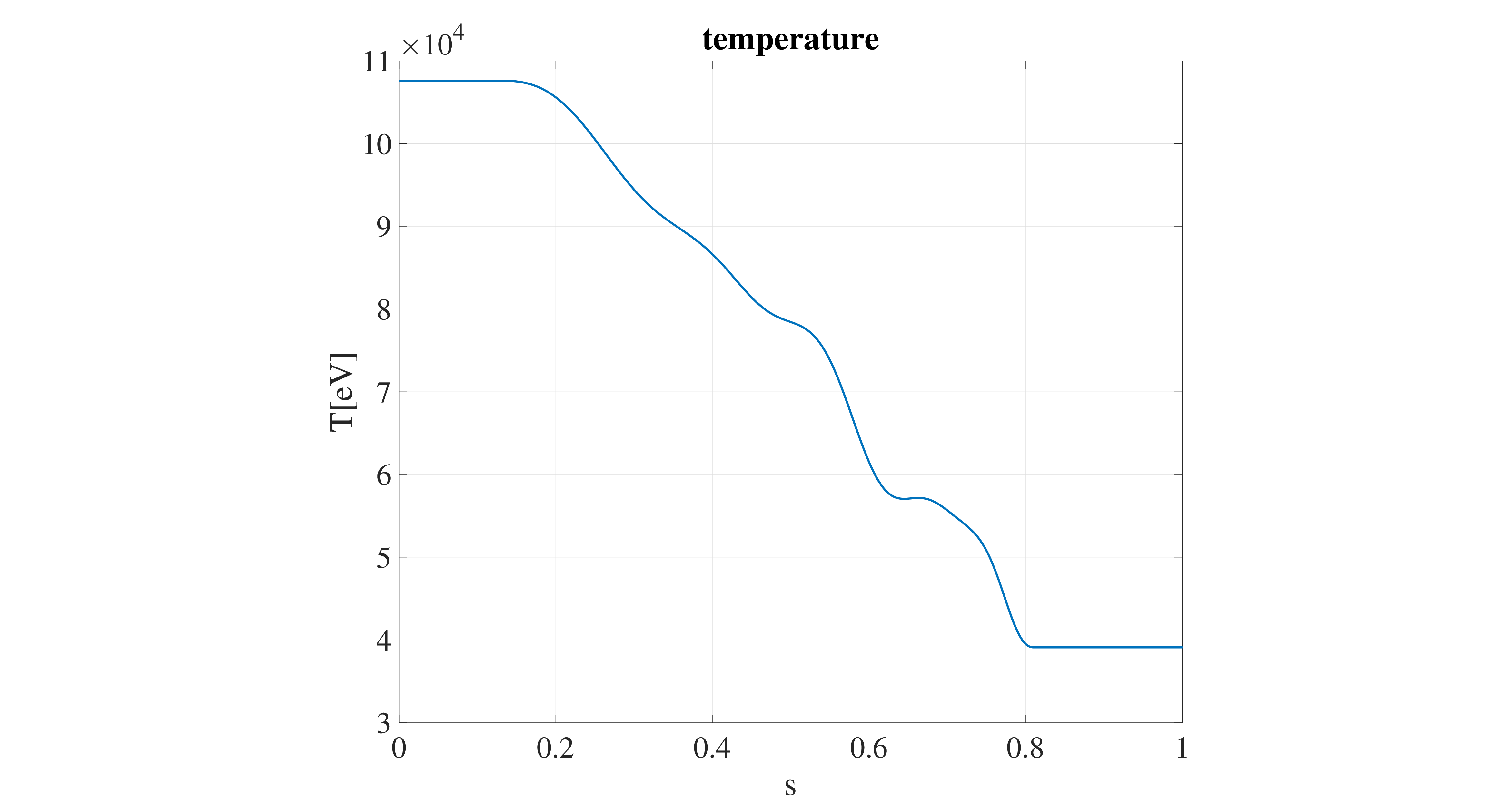}}}
     \caption{Radial profiles deduced from experimental data thanks to the TRANSP code, and taken as equilibrium initial conditions for the ORB5 simulation.(a) Density of thermal species (b)Temperature of thermal species (c) Density of energetic particles (d) Temperature of energetic particles}
     \label{fig10}
 \end{figure}

JET is a tokamak with minor radius, $a_0 = 0.95$m, major radius $R_0 = 2.88$ m, with an on axis magnetic field of $B_0 = 3.4$T. The safety factor on axis and at the edge are respective $q_0=1.74$ and $q_{edge}=4.6$. The magnetic equilibrium and safety factor profile are generated from experimental data by the CHEASE code \cite{LUTJENS1996219}, Fig \ref{fig9}. The thermal electrons, energetic species equilibrium profiles and the ion density were generated by the TRANSP code \cite{Podestà_2014} from experimental data, Fig. \ref{fig10}. However, in this study, we consider a scenario in which the electron and main ion species have the same temperature profile. A reference radial position $s_0=0.0$, is chosen from which the following physical quantities are calculated.  $\rho^*=\rho_s/a_0=3.08\cdot10^{-3}$ (with $\rho_s$ the sound larmor radius), the Alfvén speed, $V_a=7.56\cdot10^6$m/s, the sound speed, $c_s=4.8\cdot10^5$ m/s, the cyclotron frequency, $\Omega_i=1.6\cdot10^8$ rads/s and the thermal plasma beta, $\beta= 0.9\%$.

 Global linear and non-linear electromagnetic simulations were made. We use a spatial grid of $(ns,nchi, nphi) = (768, 192, 48)$ points, a time step of $dt = 3\Omega^{-1}_i$ and a number of markers of $(n_i, n_e,n_f) = (32, 128, 32)\cdot10^6$ respectively for the thermal ions (Deuterium), electrons and fast particles. A convergence test showing the validity of these simulation parameters can be found in Fig \ref{appen1}. No collisions are considered, unicity boundary condition are imposed, and heavy electrons were used to speed up the simulations ($\frac{m_e}{m_i}=0.005$). The radial domain for this simulation is $[0,1.0]$ and a filter is used such that only the $n=0,5$, toroidal modes are retained. The fraction of energetic particles for these simulations is set to $10\%$, with the $T_f/T_e=25$ at the reference location, $s_0$.
 \subsection{Linear Alfvén mode dynamics}
\begin{figure}[ht]
 \centering
\subfloat[\centering]{{\includegraphics[trim=8cm 0 8cm 0,clip,width=8cm]{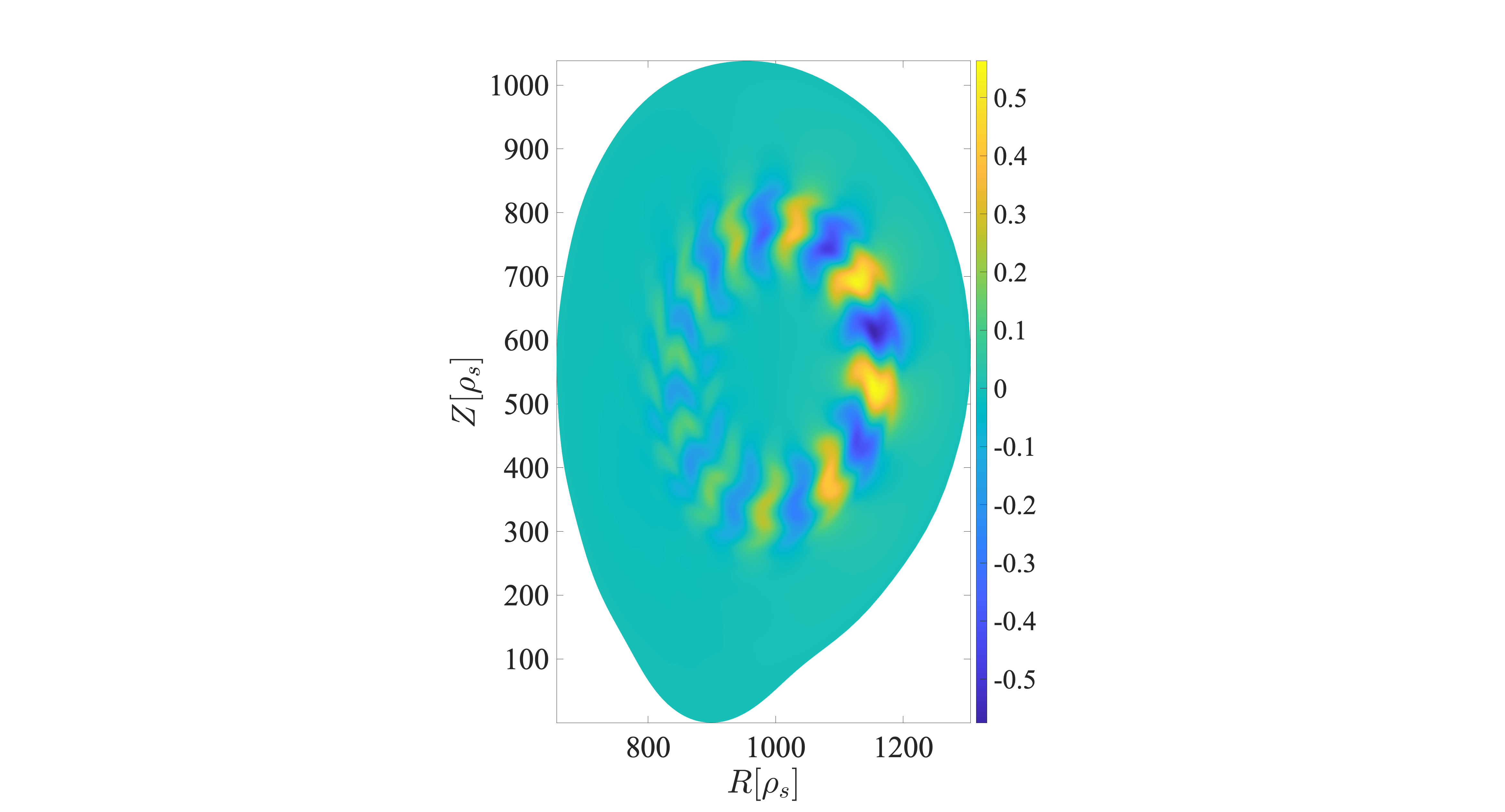}}}
\subfloat[\centering]{{\includegraphics[trim=8cm 0 8cm 0,clip,width=8cm]{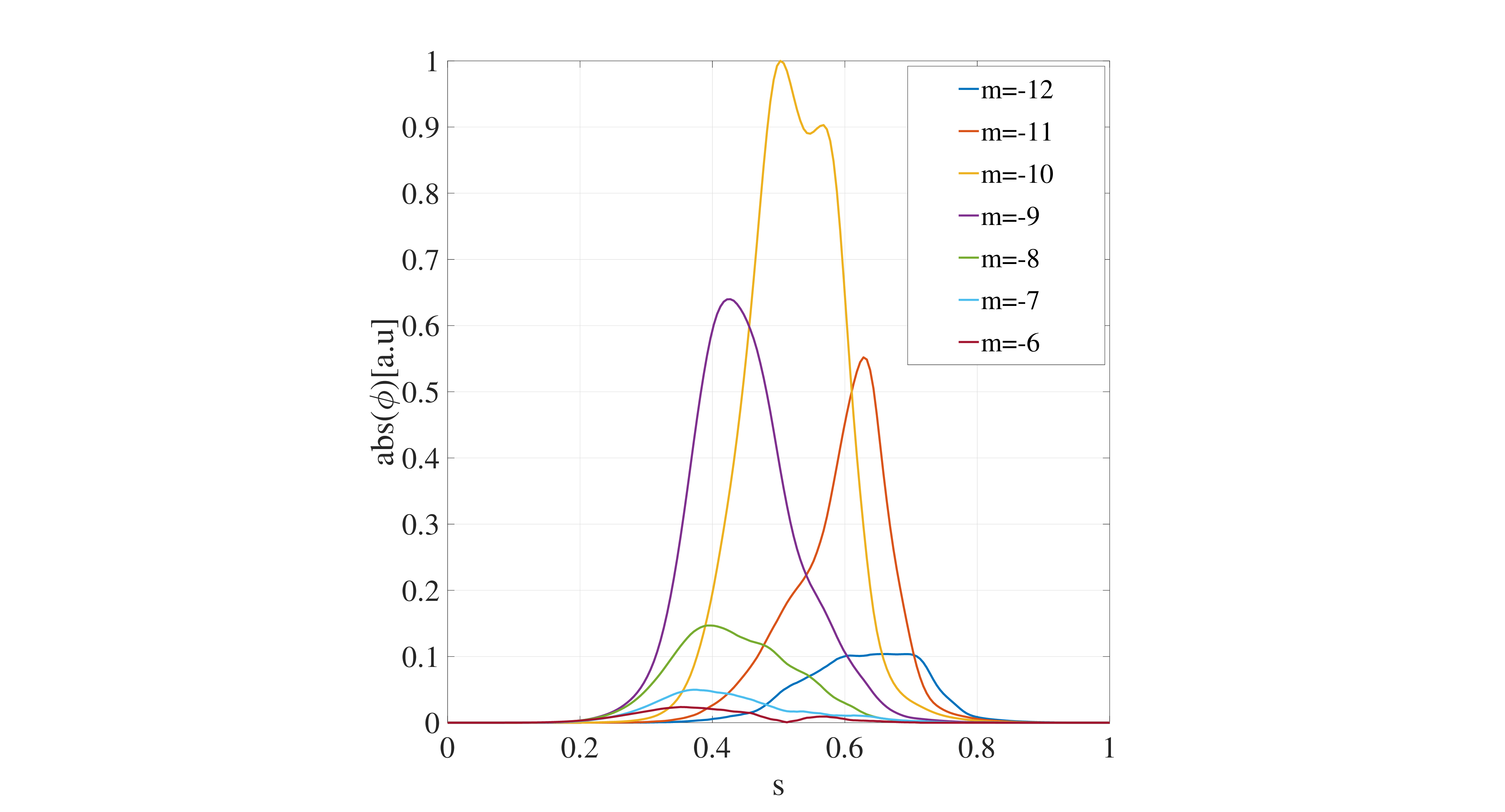}}}
     \caption{(a) Structure of TAE in the poloidal plane (b) Radial structure of individual poloidal harmonics.}
     \label{fig11}
\end{figure}

\begin{figure}[ht]
     \centering
     \includegraphics[trim=8cm 0 8cm 0,clip,width=8cm]{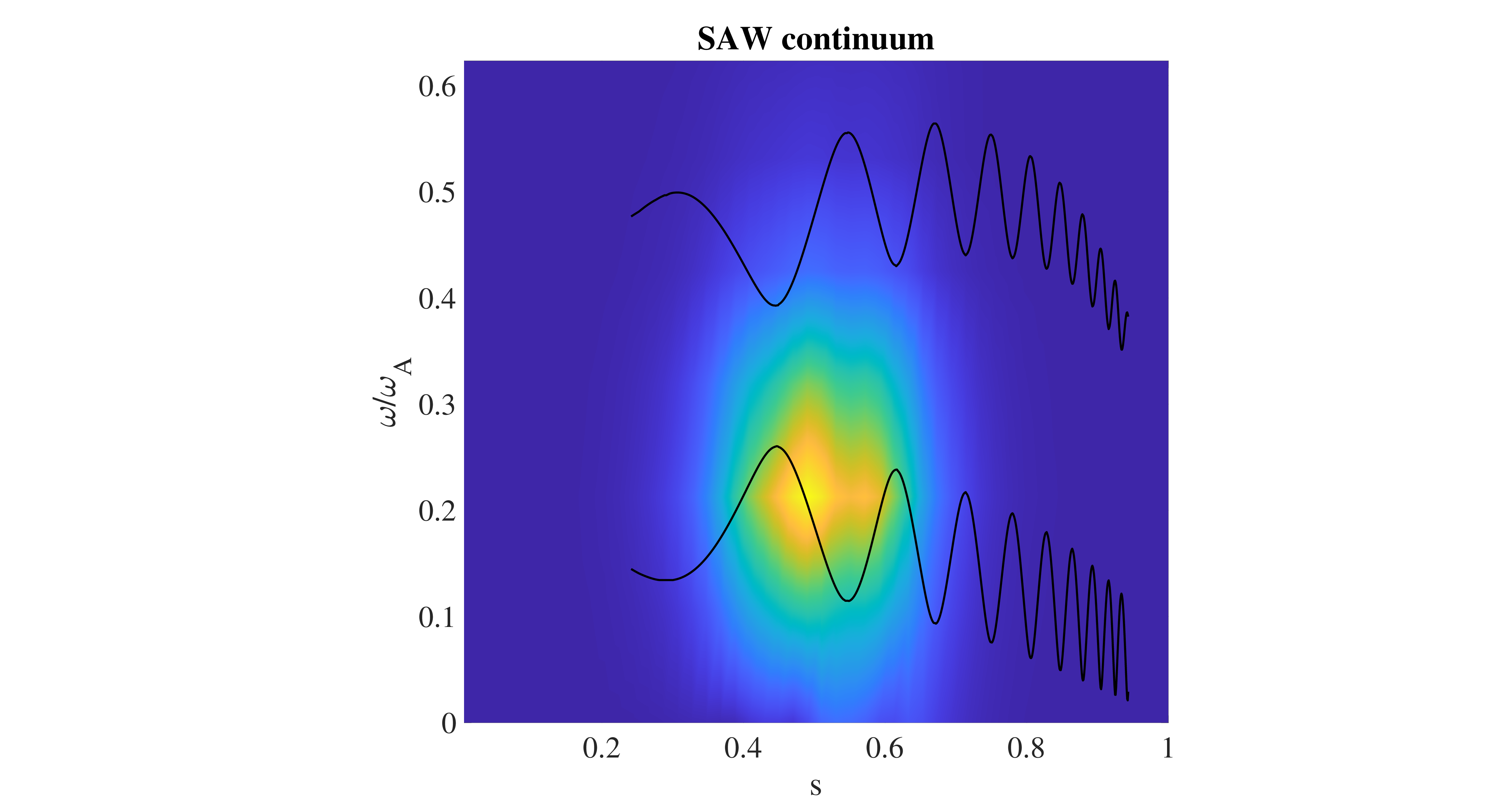}
     \caption{ Frequency spectrum showing the frequency of the excited TAEs and their radial localization. The Black curves represent the Alfvén continuum computed with the FALCON code.}
     \label{fig11b}
 \end{figure}

 An $n=5$, Alfvén mode dominated by the $m=10$, poloidal harmonic is excited by the energetic particles in the radial domain $[0.2, 0.8]$, as shown in Fig \ref{fig11} a. In the low-shear large-aspect ratio limit, the eigenmode amplitude is not strongly localized around half-rational surfaces, and individual poloidal harmonics are active across an extended radial region. The strong response at $s=0.5$ results from the coupling between the $m=(9,10)$ poloidal harmonics, in contrast to the response at $s=0.58$ which is due to the coupling between the $m=(10,11)$. If we decompose the scalar potential into its poloidal harmonics and plot the radial structure of these harmonics at a particular time in the linear growth phase of the modes, we can clearly observe the dominant poloidal harmonics and deduce the positions of maximum coupling, Fig \ref{fig11} b. The localization of this mode as well as the dominant poloidal harmonics are in accordance with theory \cite{berk}.  
 The excited mode resulting from the coupling between different poloidal harmonics at different radial locations, has growth rate and frequency, which are respectively, $\gamma=0.00153\Omega_i$, with a frequency of $\omega\approx 100$kHz. This frequency fall in the TAE range (which lies between $60$kHz to $200$kHz). We can check that the excited modes are really TAEs by comparing their radial localization and frequencies to the Shear Alfvén continuum spectrum (SAW) corresponding to the equilibrium in question. The FALCON code \cite{falessi} was used to calculate the SAW spectrum and its result is plotted on the frequency spectrum measured from the ORB5 simulation, Fig \ref{fig11b}. We observe from this plot that the frequencies of the excited modes fall in the TAE gap. So the excited Alfvén modes, are identified as TAEs. This plot also shows that the strongest TAE results from the coupling between the poloidal harmonics $m=(9,10)$ at $s=0.5$. It should be noted that due to the high fraction of energetic particles used in this simulation, the mode growth rate constitutes a reasonable fraction of the frequency, i.e. the fast particles are starting to have an effect on mode dispersion. This explains why the mode touches the continuum.

\subsection{Non-linear Alfvén dynamics }
\begin{figure}[ht]
    \centering
    \includegraphics[trim=8cm 0 8cm 0,clip,width=8cm]{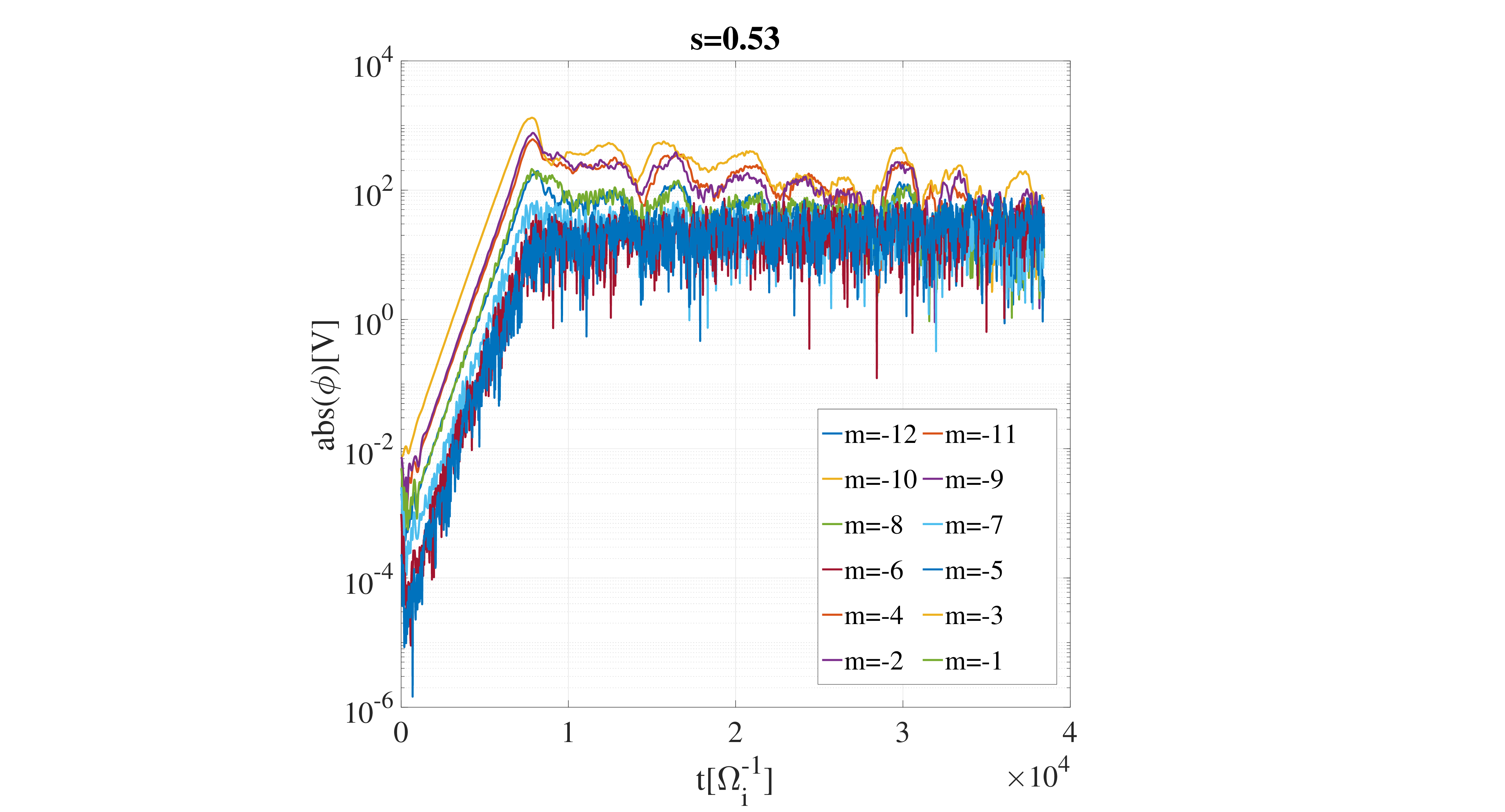}
    \caption{Time evolution of poloidal modes at mid-radius.}
    \label{fig:enter-label}
\end{figure}

\begin{figure}[ht]
 \centering
\subfloat[\centering]{{\includegraphics[trim=8cm 0 8cm 0,clip,width=8cm]{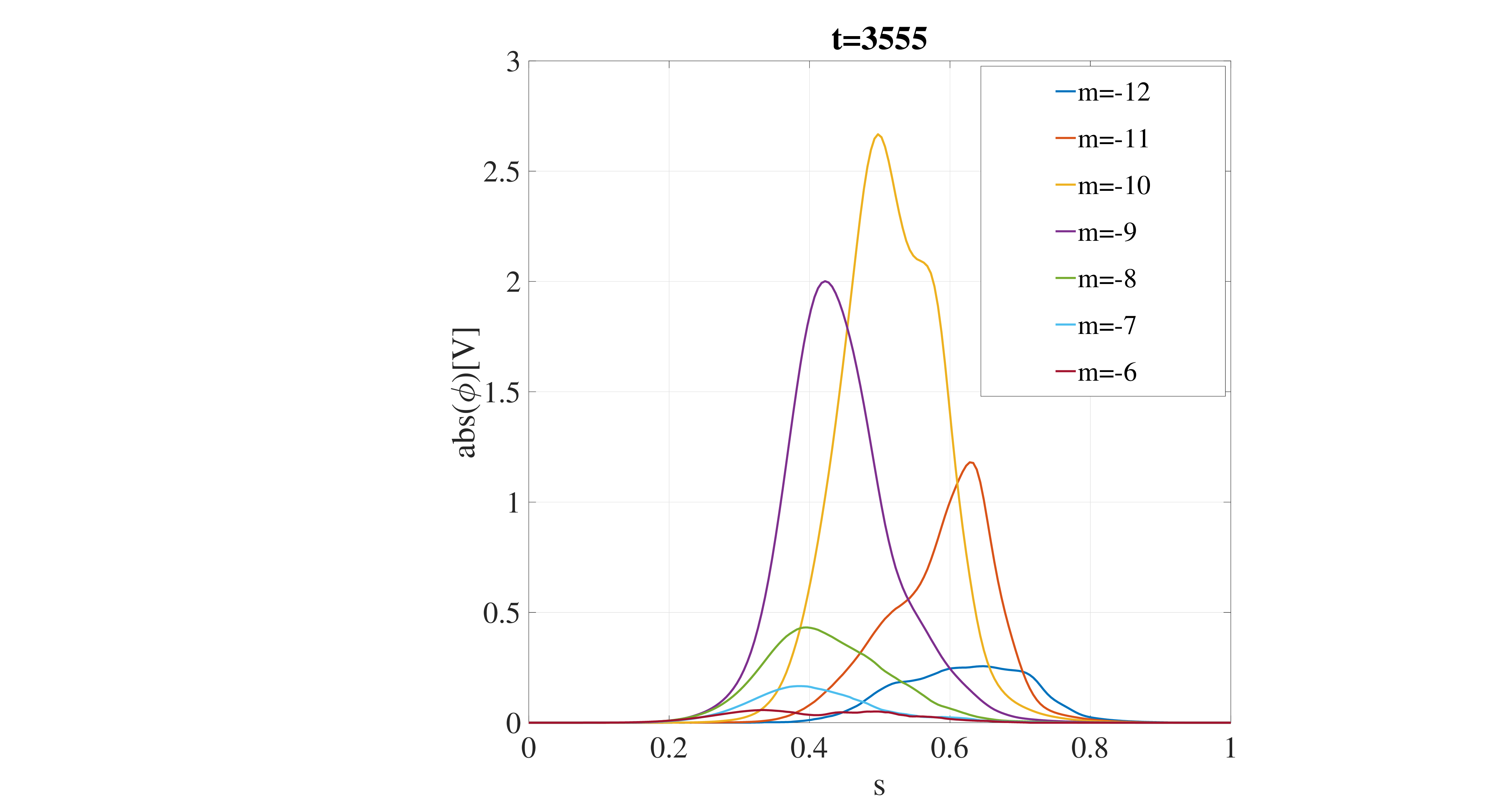}}}
\subfloat[\centering]{{\includegraphics[trim=8cm 0 8cm 0,clip,width=8cm]{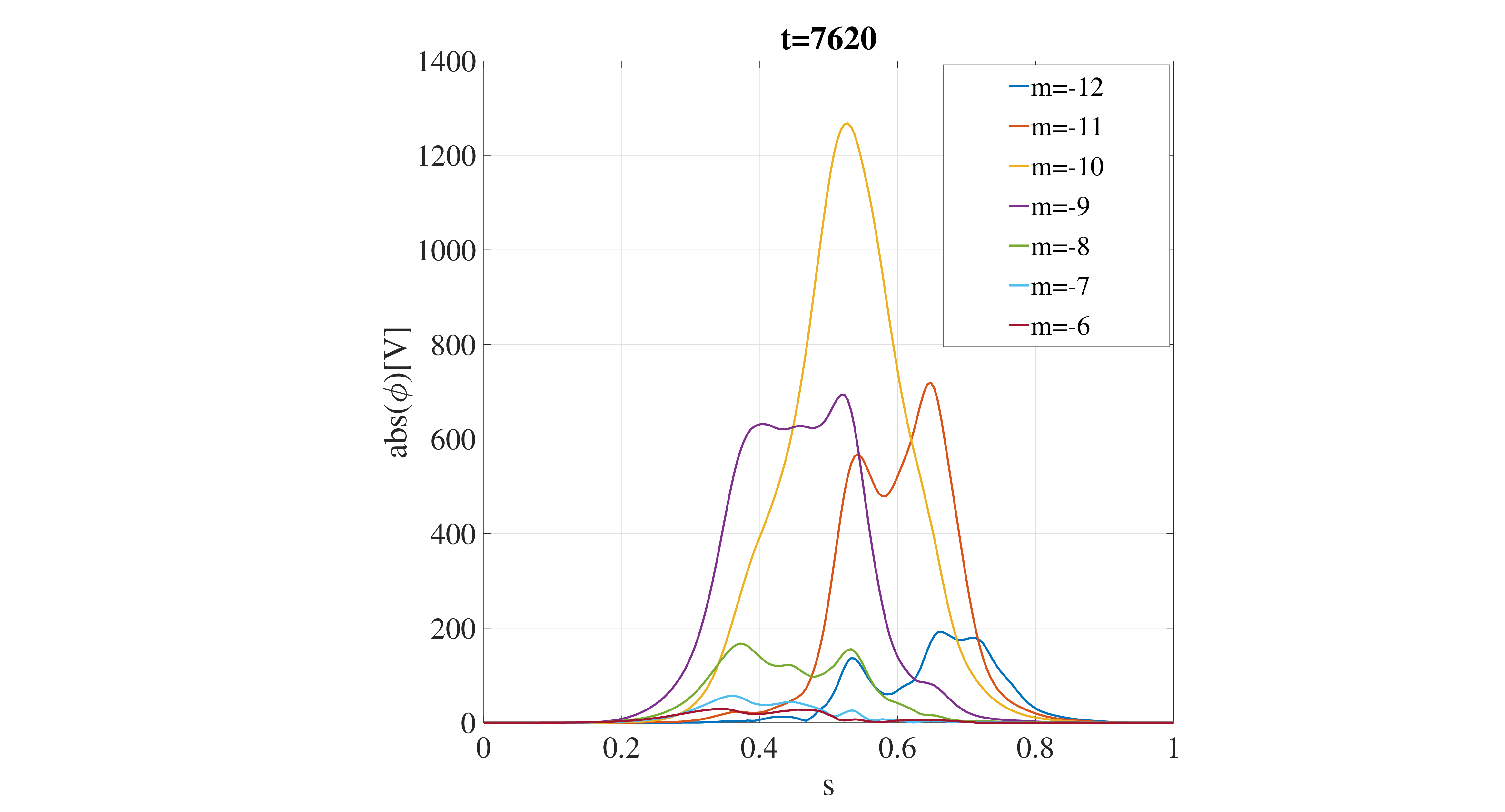}}}
\newline
\subfloat[\centering]{{\includegraphics[trim=8cm 0 8cm 0,clip,width=8cm]{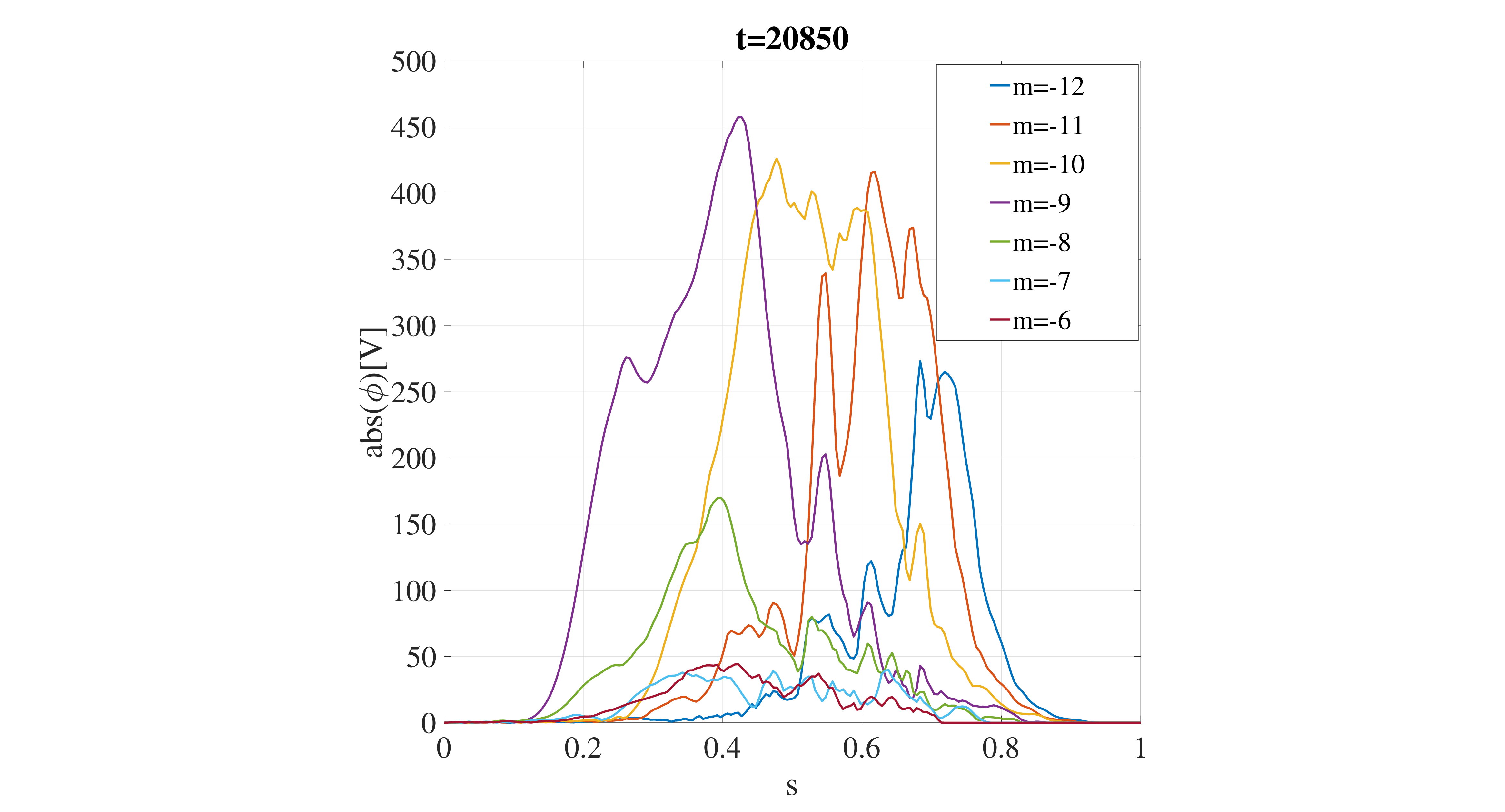}}}
\subfloat[\centering]{{\includegraphics[trim=8cm 0 8cm 0,clip,width=8cm]{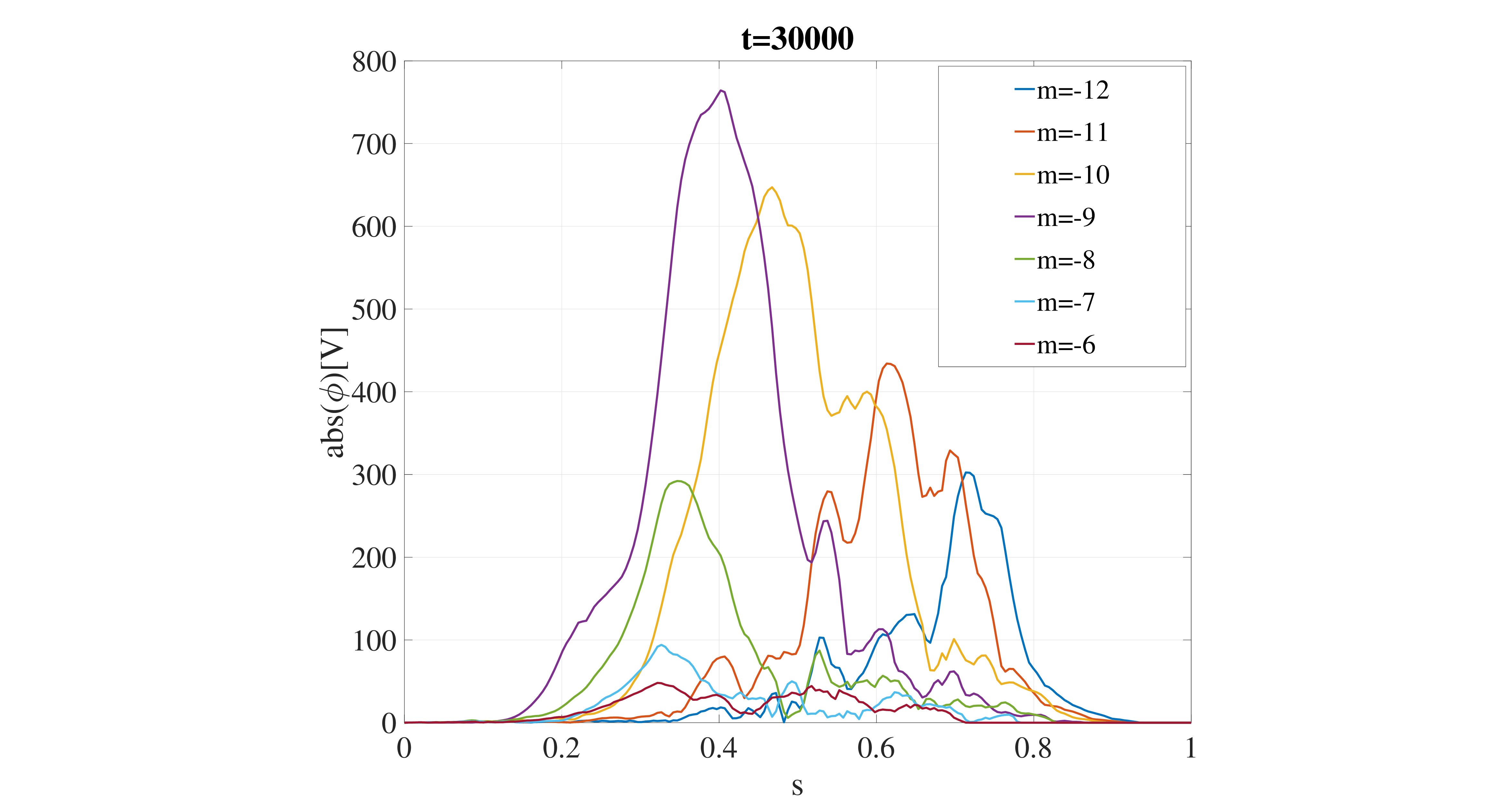}}}
     \caption{Mode structure of excited mode (a) Linear phase (b)Early saturation phase (c) Early nonlinear phase (d) Deep nonlinear phase.}
     \label{fig12a}
\end{figure}

A filter is applied in the non-linear simulation such that only the $n=0,5$, are retained in the simulation. Fig \ref{fig:enter-label}, shows the time evolution of the of different poloidal harmonics. The linear phase is completely dominated by $m=10$ harmonic. The excited modes in the linear phase are localizes in the radial domain $[0.2,0.8]$, Fig \ref{fig12a}. In nonlinear phase, however, the dominant linear harmonic saturates at amplitudes of the same order as the $m=9,11$. These three modes dominates the nonlinear phase. The nonlinear phase is also characterized by an increase in the coupling between the dominant poloidal harmonics and other poloidal harmonics such as the the $m=8,12$, Fig \ref{fig12a} c and d. The non-linear phase also leads to coupling between different toroidal harmonics. The couple between the $n=0,5$ will be discussed in the next subsection. This complex coupling that occurs in the nonlinear phase leads to a modification of the mode structure of the excited mode and a radial spreading of the instability. Fig \ref{fig12a}, shows how the mode structure of the TAE changes over time.
\subsection{Non-linear excitation of Zonal flows by TAEs}
\begin{figure}[ht]
 \centering
\subfloat[\centering]{{\includegraphics[trim=8cm 0 8cm 0,clip,width=8cm]{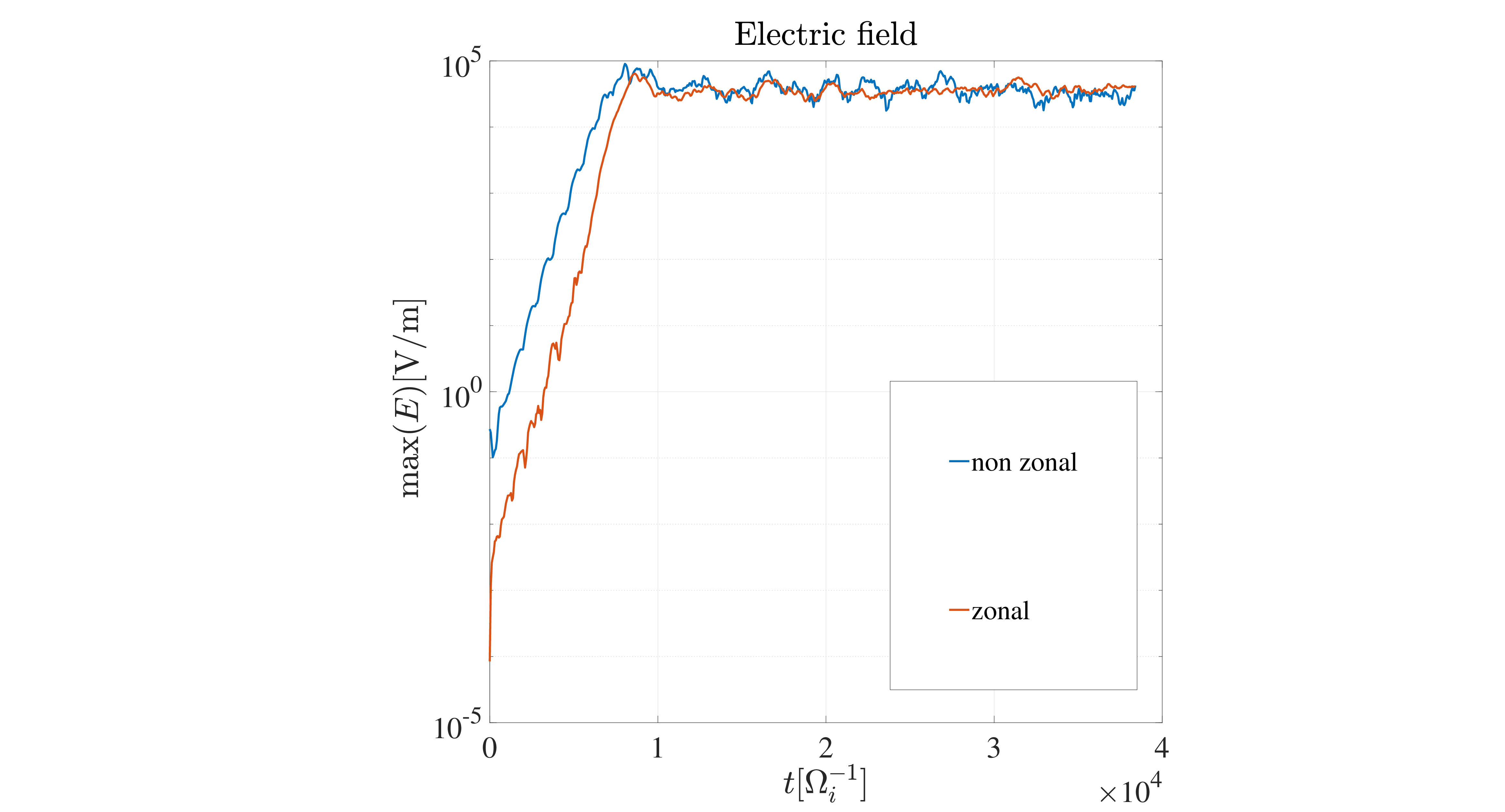}}}
\subfloat[\centering]{{\includegraphics[trim=8cm 0 8cm 0,clip,width=8cm]{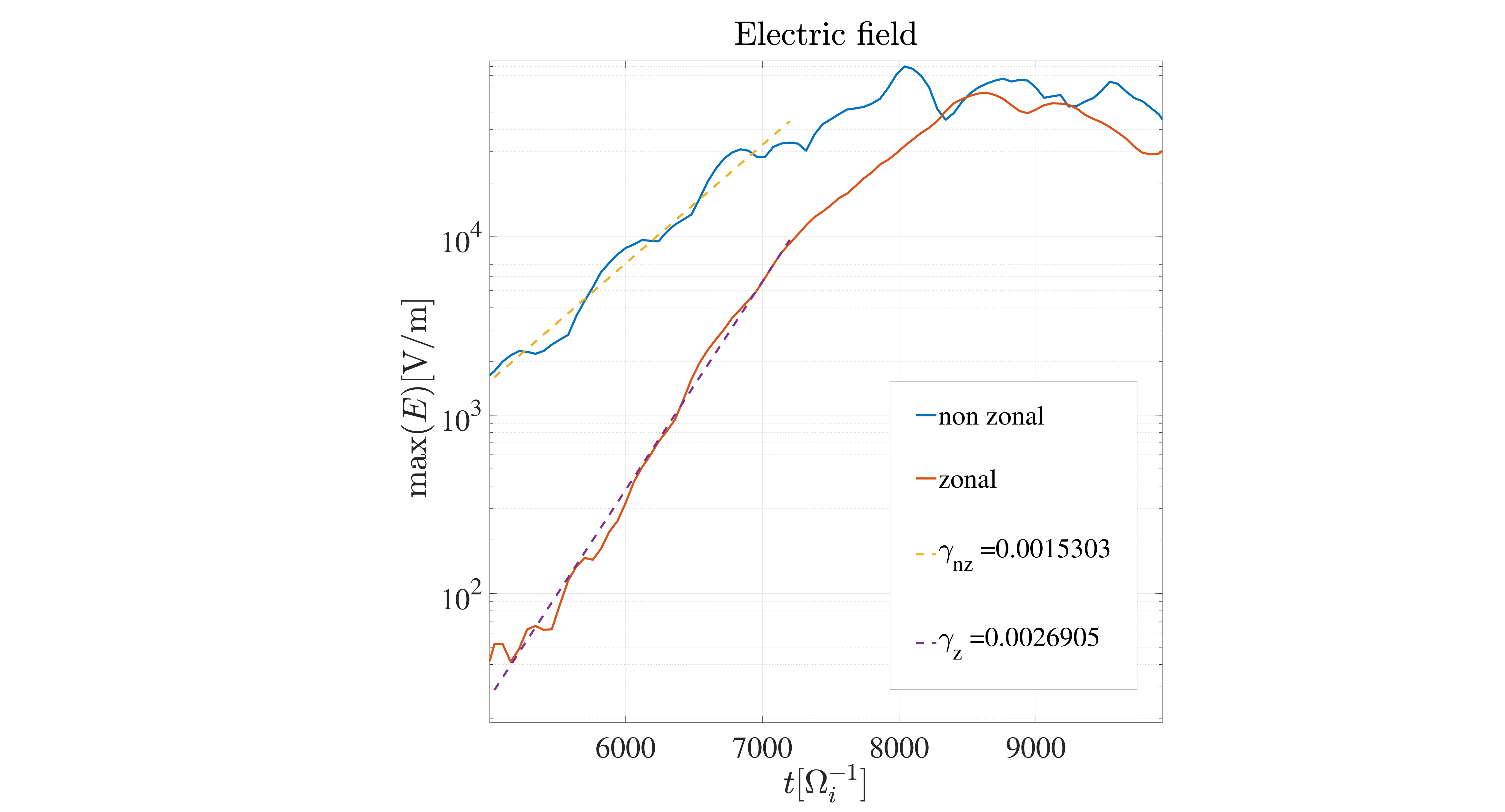}}}
\newline
\subfloat[\centering]{{\includegraphics[trim=8cm 0 8cm 0,clip,width=8cm]{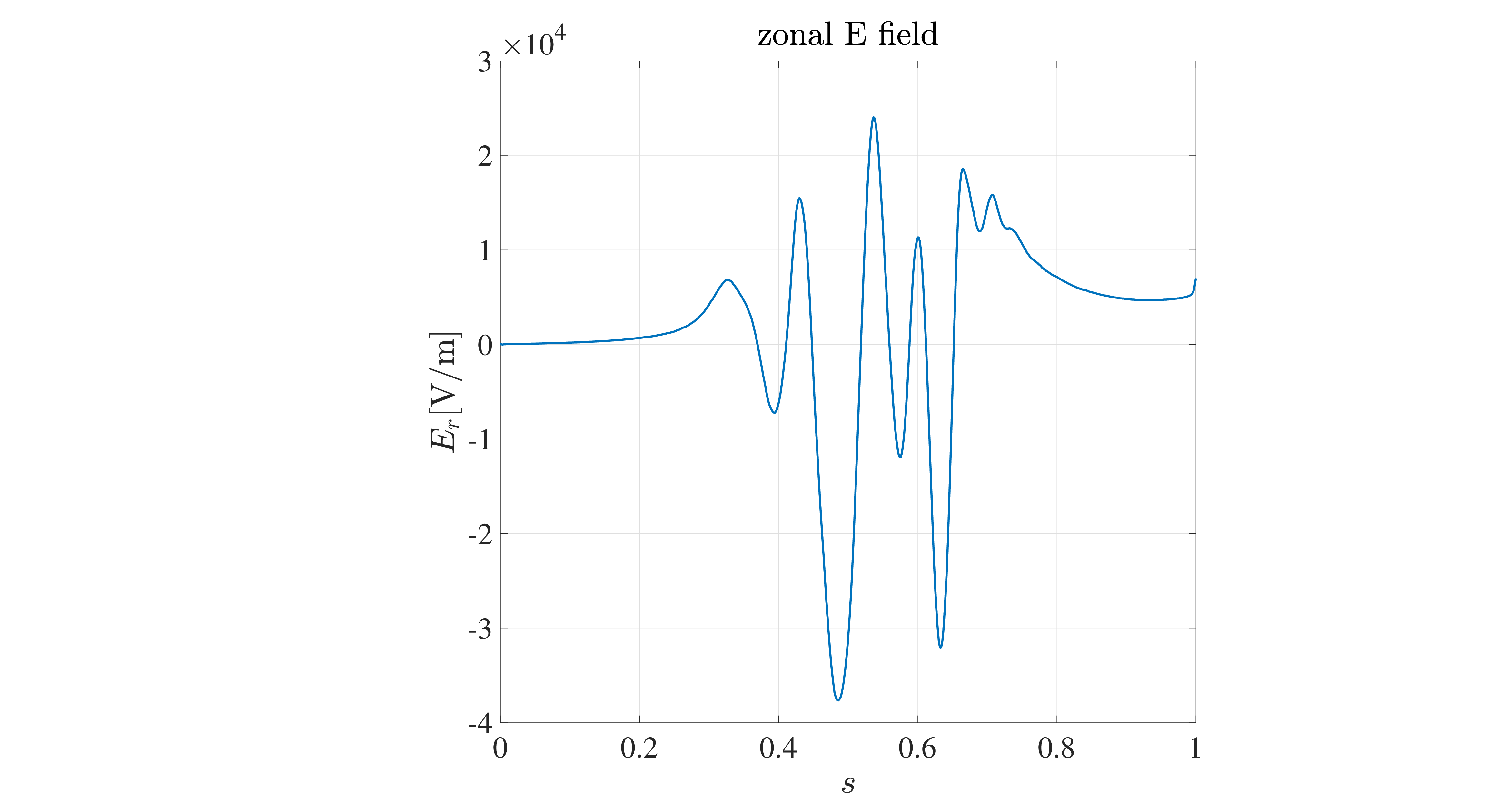}}}
\subfloat[\centering]{{\includegraphics[trim=8cm 0 8cm 0,clip,width=8cm]{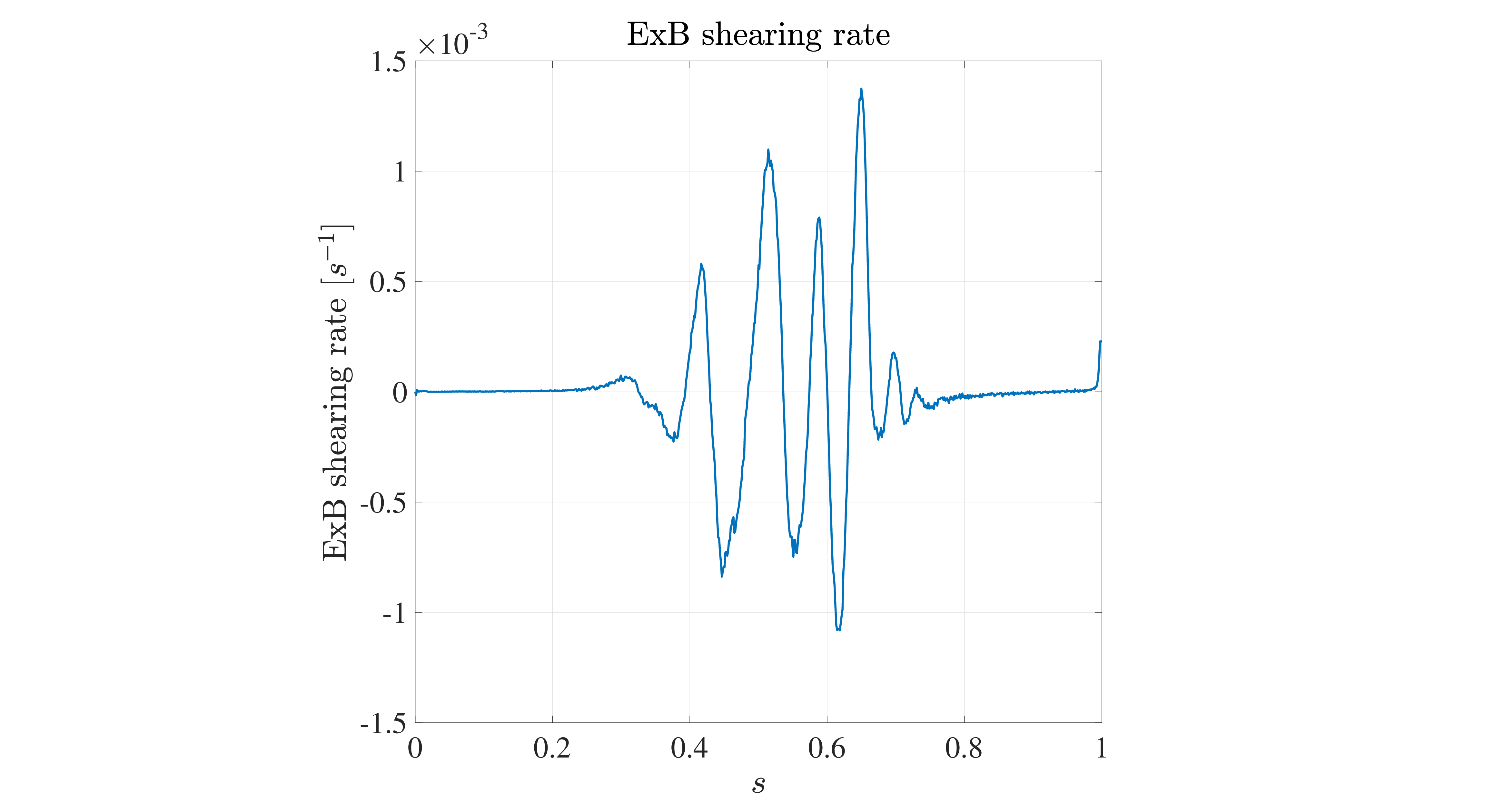}}}
     \caption{(a) Time evolution of the electric field showing the excitation of ZF by TAE (b)Zoom of early nonlinear phase with ZF growing nearly two times faster than TAE (c) Zonal flow radial structure in the saturation phase. The fine radial structure of the ZF can be observed (d) The zonal flow ExB shearing rate in the saturation phase.}
     \label{fig12}
\end{figure}

The nonlinear phase is also characterized by the coupling between the $n=0,5$, toroidal harmonics. This coupling leads to excitation of the zonal flows ($n=0$) by the $n=5$ TAE in the early nonlinear phase, Fig \ref{fig12}a
. According to Qiu \cite{qiu}, in the presence of energetic particles, Alfvén modes can drive zonal flows via forced-driven excitation. This excitation mechanism is characterized by the zonal flows growing at a rate that is approximately twice that of the Alfvén mode driving it. In Fig \ref{fig12} b, we can observe that in early nonlinear phase, the zonal flow driven by the TAE has a growth rate that is approximately twice that of the driving TAE. These zonal flows are therefore forced driven. A frequency analysis of this zonal flow shows that it is a zero frequency zonal flows. The TAE and the zonal flow saturate respectively with electric field amplitudes of $90$kV/m and $63.5$kV/m. The amplitude of the zonal electric field is consistent with the magnitude of the same quantity measured in other JET shots, e.g. Fig. 7 in \cite{Crombé_2009}. The radial structure of the excited zonal flow and its associated shear is shown in Figs. \ref{fig12} c and d, respectively. This zonal flow has a fine radial structure as compared to the driving TAE radial structure. This fine radial structure of zonal flows excited by Alfvén modes was also predicted in theory \cite{qiu}. With these measurements, we can configure an "antenna" in ORB5 to emulate this zonal flow, in a simulation in which we aim to measure its impact, $-$see next section.

\section{Effects of forced-driven zonal flows on ITG dynamics} 
\subsection{Dynamics in the absence of zonal flows}
\begin{figure}[ht]
     \centering
\subfloat[\centering]{{\includegraphics[trim=8cm 0 8cm 0,clip,width=8cm]{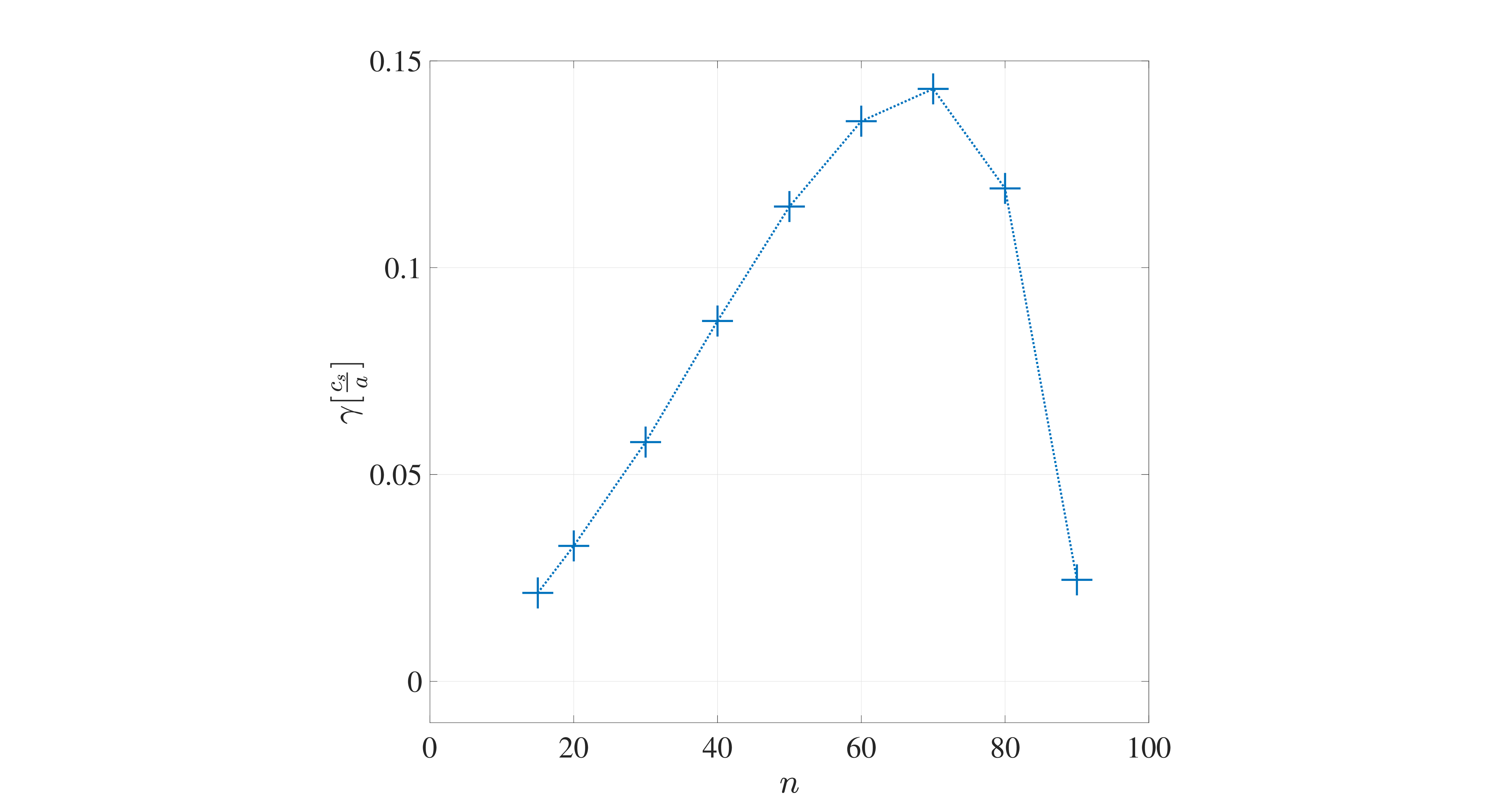}}}
\subfloat[\centering]{{\includegraphics[trim=8cm 0 8cm 0,clip,width=8cm]{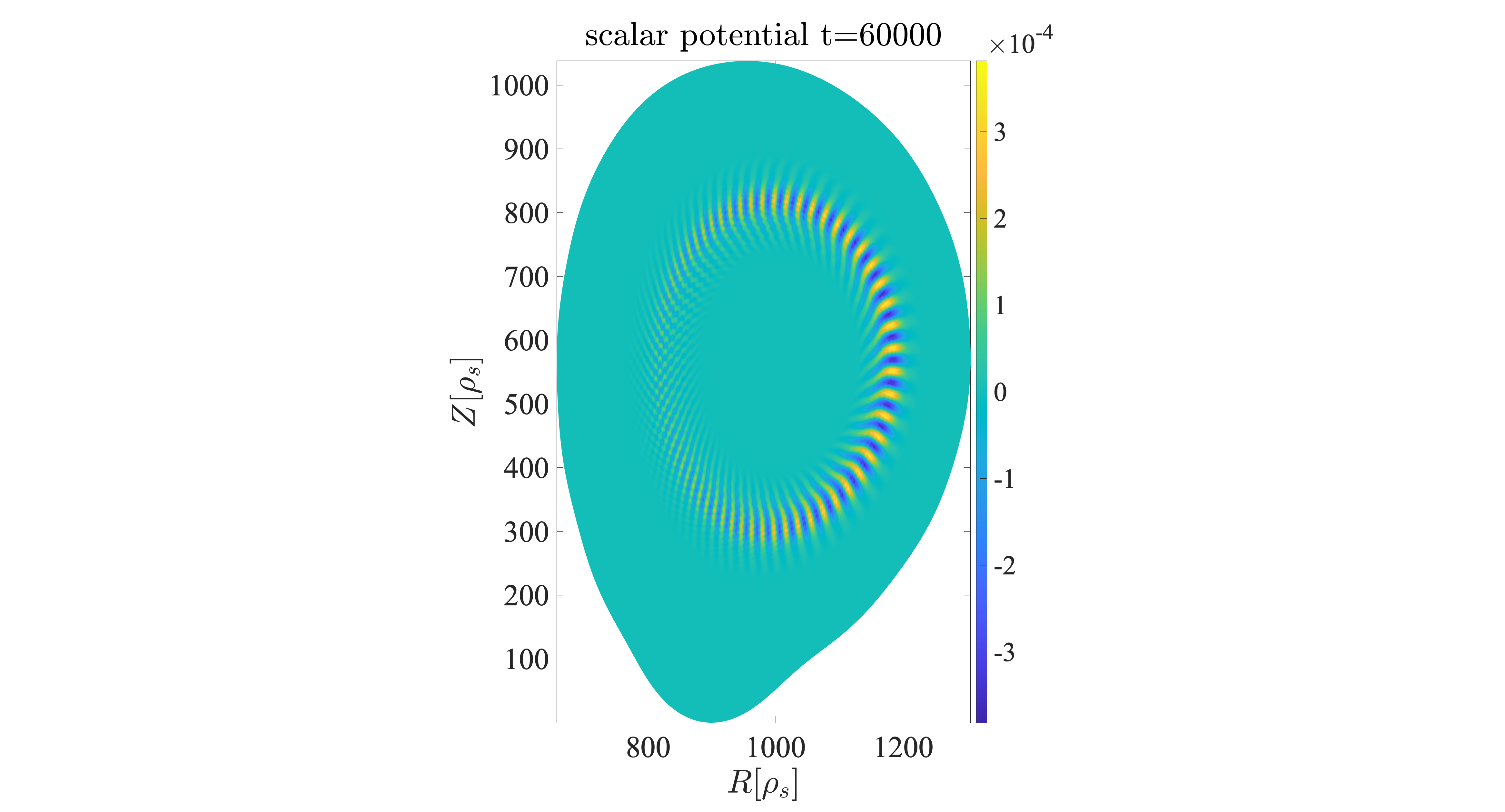}}}
     \caption{(a) ITG growth rate versus toroidal mode number (b) $n=20$ ITG structure in the poloidal plane }
     \label{fig13}
 \end{figure}

In order to measure the effects of the forced driven zonal flows on ITG instability, in this experimental scenario, we first have to study the dynamics of this instability in the absence of any source of zonal flows. We consider, the same simulation parameters as in the previous case. Electrons in this case are adiabatic, and we don't include energetic particles, and we consider only electrostatic fluctuations. A scan in toroidal mode numbers is performed and the corresponding linear ITG growth rates are measured. The fastest growing ITG mode has a toroidal mode number of $n=70$, Fig. \ref{fig13} a. The ITG structure in the poloidal plane is shown in Fig. \ref{fig13} b, with the excited mode being most unstable in the low field side of the device. These modes are excited around the mid-radius in the low field side of the device, where there is a peak in the ion temperature gradient.
 \subsection{The ORB5 antenna module}
  The antenna module, first implemented in ORB5 by Ohana and co-workers \cite{noe,Sadr}, was designed to study the impact of specific plasma modes on the dynamics of plasma instabilities or turbulence. The antenna acts like an external source whose field can emulate a specific plasma mode given the frequency ($\omega$), toroidal ($n$) and poloidal ($m$) mode numbers, the radial structure of the mode($h(s)$) and the phase offset ($\Phi$). However, the antenna and plasma field are considered separately, so that in linear simulations, only the plasma fields are linearized keeping the antenna fields in the particle's characteristics. This configuration makes it possible to use the antenna both for linear and non-linear studies. In its simplest form, the antenna field, $F_{ant}$, can be written as;
 \begin{equation}
     F_{ant}=h(s)Re\left[\sum_{j=1}^NA_je^{i\left(m_j\theta+n_j\varphi+\Phi_j\right)}\left(\alpha+\beta e^{i\omega t}\right)\right]
 \end{equation}
 Where $A_j$, is the Fourier mode coefficient, $\alpha$ and $\beta$ are respectively the coefficients of the static and oscillating components. The quantity $F_{ant}$ can be the electrostatic potential $\phi$ and/or the parallel vector potential $A_{||}$. In this work, we used the electrostatic antenna component only.
  For this present study, the zonal flows described in the section above can be emulated  by the ORB5 antenna module, by setting, $m_j=n_j=\omega=\Phi_j=0$, while the scalar potentials corresponding  to zonal electric fields structures shown in Fig. \ref{fig12} c is loaded into $h(s)$.
 \subsection{ITG dynamics in the presence of forced driven zonal flows}
\begin{figure}[ht]
\centering
\subfloat[\centering]{{\includegraphics[trim=8cm 0 8cm 0,clip,width=8cm]{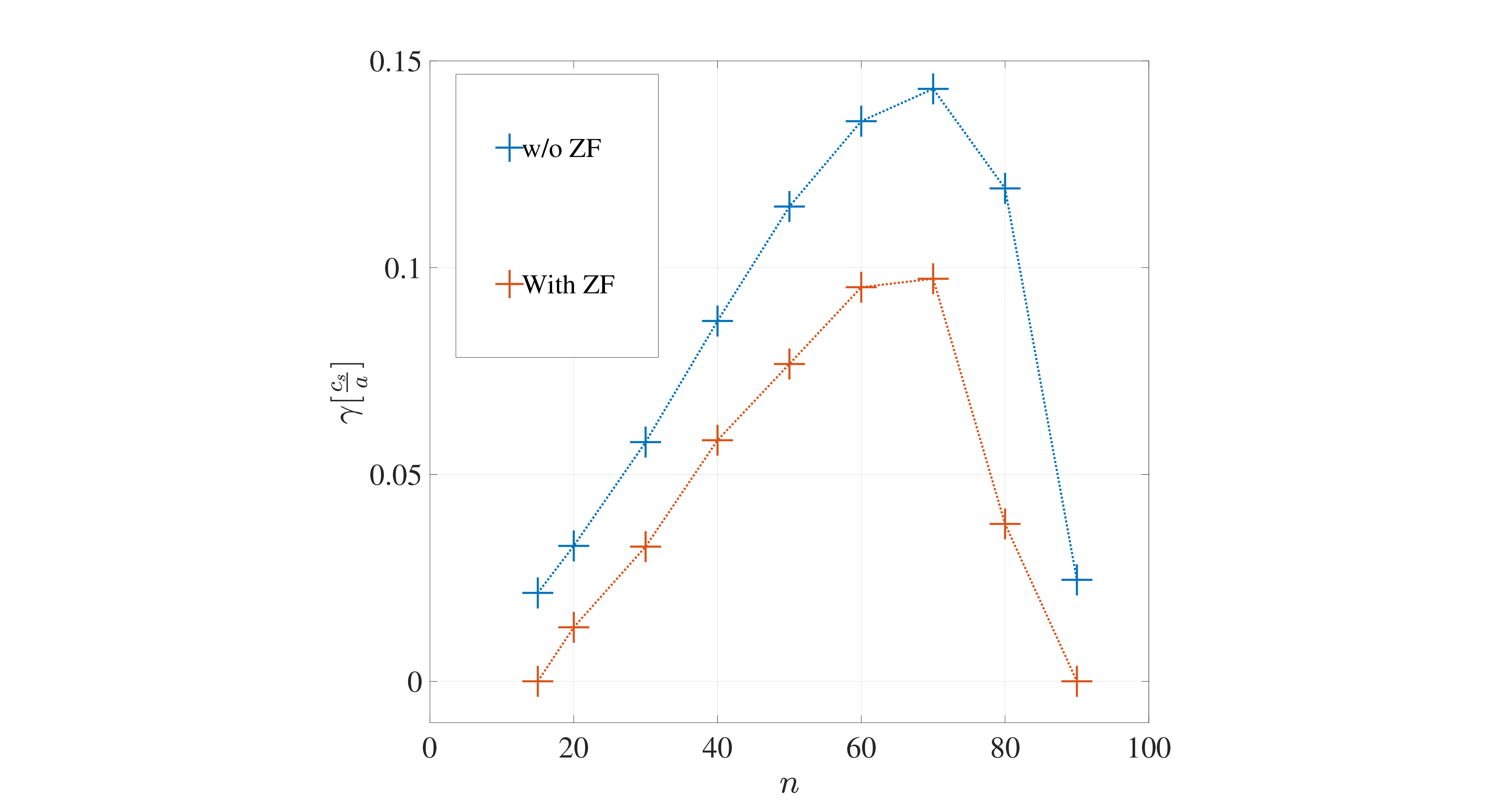}}}
\subfloat[\centering]{{\includegraphics[trim=8cm 0 8cm 0,clip,width=8cm]{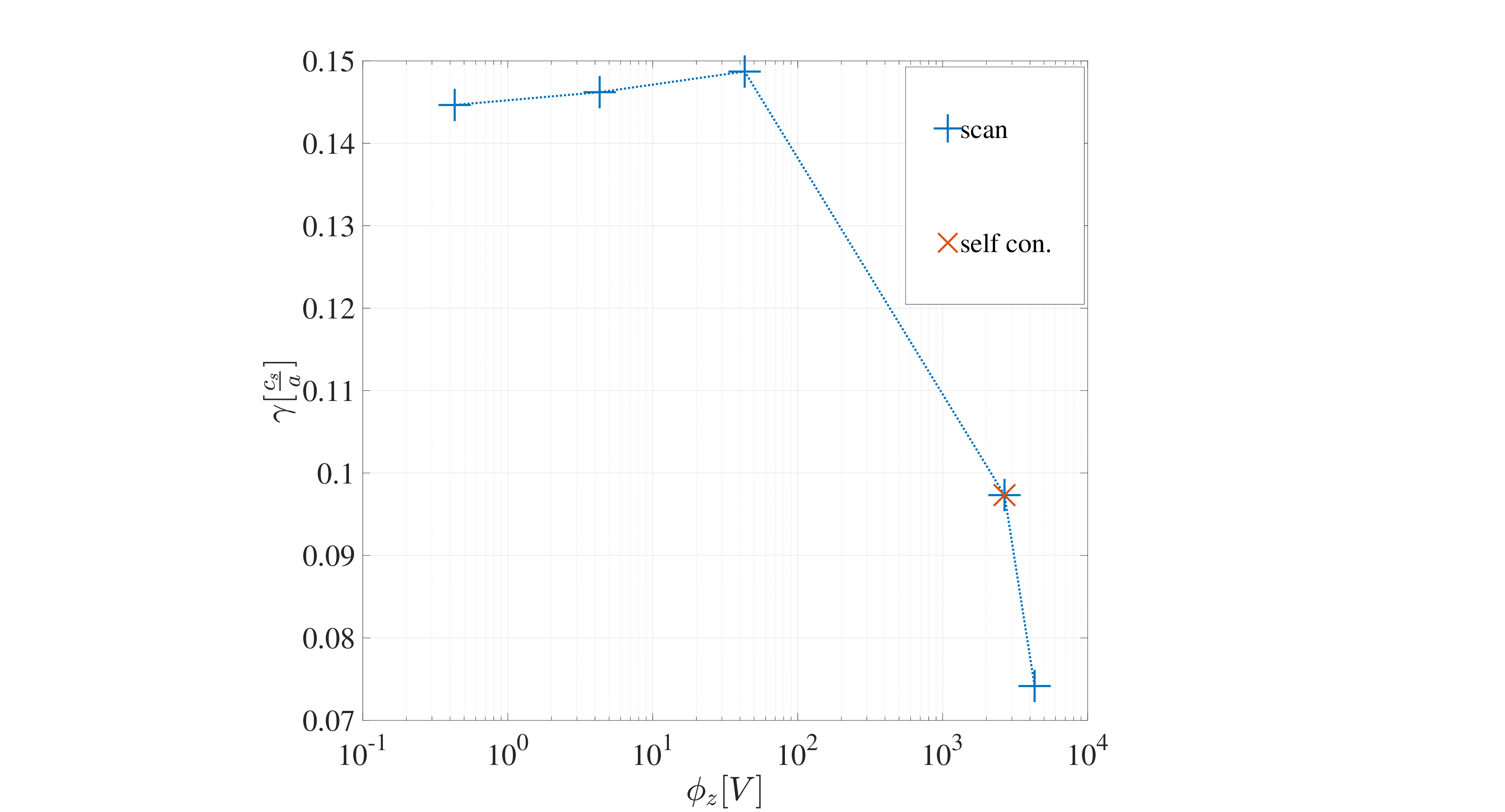}}}

     \caption{(a)ITG growth rate versus toroidal mode number  without and with forced-driven zonal flows, for the parameters and equilibrium configuration discussed in Sec.\ref{equi}. (b) Zonal flow amplitude scan, showing stronger stabilization for higher amplitudes.}
     \label{fig14}
 \end{figure}

 In this section, we use the ORB5 antenna module described earlier to model the zonal flows forced-driven by the $n=5$ TAE. The excited zonal flow is a ZFZF, with $n=m=0$. The radial structure of the scalar potential from which the zonal electric field is derived is loaded into the antenna module. We perform single toroidal mode ITG simulations and scan over different toroidal mode numbers, and the corresponding ITG growth rates in the presence of the ORB5 antenna emulating the forced-driven zonal flow are measured. This zonal flow produces a significant mitigation of the ITG instability, measured by the large reduction of the growth rate of the most unstable mode that dominates the linear dynamics. All toroidal mode numbers are strongly stabilized in the presence of forced-driven zonal flows and the range of unstable modes is correspondingly reduced, Fig. \ref{fig14} a. This result shows that the mitigation of ITG instability by forced-driven zonal flows is significant in experimentally relevant  conditions. Therefore, it can serve as an important indirect part of turbulence mitigation by energetic particle via excitation of forced driven zonal flows by Alfvén modes. These simulations have shown the effectiveness of such a process.
 In different experimental scenarios, we can expect significant differences in energetic particles population and temperature. These differences will generally results in a change in the Alfvén modes and zonal flows saturation levels. It is therefore important to study how the zonal flow amplitude affects the stabilization of ITG modes. To this end, we focus on the most unstable linear mode i.e. $n=70$, and perform a zonal flow amplitude scan, while keeping the radial structure constant. We observe that the mitigation of the ITG modes doesn't occur starting from a certain threshold value of amplitude. The ITG dynamics rather changes with the zonal flow amplitude, with the mitigation of the instability increasing with the zonal flow amplitude , Fig \ref{fig14} b. The zonal flow amplitude is found to be inversely proportional to the ITG growth rate. This inverse relations have also been reported in theory\cite{ningfei}. The stabilizing effect of radial electric field on ITGs has also been reported in \cite{Allfrey_2002}.
\section{Conclusion}
In this work, we investigated the indirect interaction mechanism of turbulence and energetic particles, via forced-driven zonal flows. These forced-driven zonal flows are driven by Alfvén modes excited by energetic particles. Biancalani and co-workers \cite{alessandro}, proposed a mechanism in which the forced driven zonal flows excited by the Alfvén modes can be used to mitigate background turbulence. We designed a numerical experiment to test this conjecture by running linear and non-linear gyro kinetic simulation with the particle in cell code ORB5. The approach we used consisted in running two types of simulations. A nonlinear simulation in which zonal flows are forced-driven by Alfvén modes excited by energetic particles. The excited zonal flows are measured, and its parameters are used to configure the antenna module of ORB5, which is then used to emulate this zonal flow in a linear ITGs simulations in order to study its effects on the ITG dynamics. The study was carried out using magnetic equilibrium and species profiles generated from experimental data from the JET shot 92416 Afterglow experiment.

We found in the Alfvén mode simulations that the excited $n=5$ mode was a TAE as its frequency falls in the TAE gap of continuum spectrum corresponding to this equilibrium. This TAE nonlinear excited a zero frequency zonal flow that had a growth rate that was twice that of the TAE, an evidence of forced-driven excitation \cite{qiu}. The quantitative features of the excited zonal flow have been identified, and its parameters were used to configure an antenna in ORB5. Running linear ITG simulation with the ORB5 antenna emulating the forced-driven zonal flows, we found that ITG dynamics was significantly affected, leading to a significant reduction of the growth rate of these modes. These results show that in experimentally relevant conditions, the zonal flows, forced driven by Alfvén modes due to their excitation by energetic particles, can significantly impact the ITG instability dynamics. Hence, the indirect interaction of energetic particles with turbulence through the excitation of forced driven zonal flows, can be an important indirect path of turbulence mitigation. The present ITG study is restricted to the electrostatic, adiabatic electron model, and further works will examine the effect of forced-driven ZFs on other types of micro-instabilities, using kinetic electrons.
\section*{Acknowledgement}
This work was partially supported by the “Lorraine Université d’Excellence”  Doctorate  fundings  (project R01PKJUX-PHD21)  belonging to the Initiative "I-SITE LUE". Part of this work has been carried out within the framework of the EUROfusion Consortium, funded by the European Union via the Euratom Research and Training Programme (Grant Agreement No.$101052200$ EUROfusion). Views and opinions expressed here are however those of the authors only and do not necessarily reflect those of the European Union or the European Commission. Neither the European Union nor the European Commission can be held responsible for them. Numerical calculations for this work have been partially performed on the cluster Explor of the "Maison de la simulation Lorraine" (We are grateful to the partial time allocation under the project No. 2019M4XXX0978), and on the MARCONI FUSION HPC system at CINECA.
\begin{appendices}
\section{Convergence test of nonlinear Alfvén mode simulation}
\begin{figure}[htbp]
\centering
\includegraphics[trim=5cm 0 5cm 0,clip,width=0.5\linewidth]{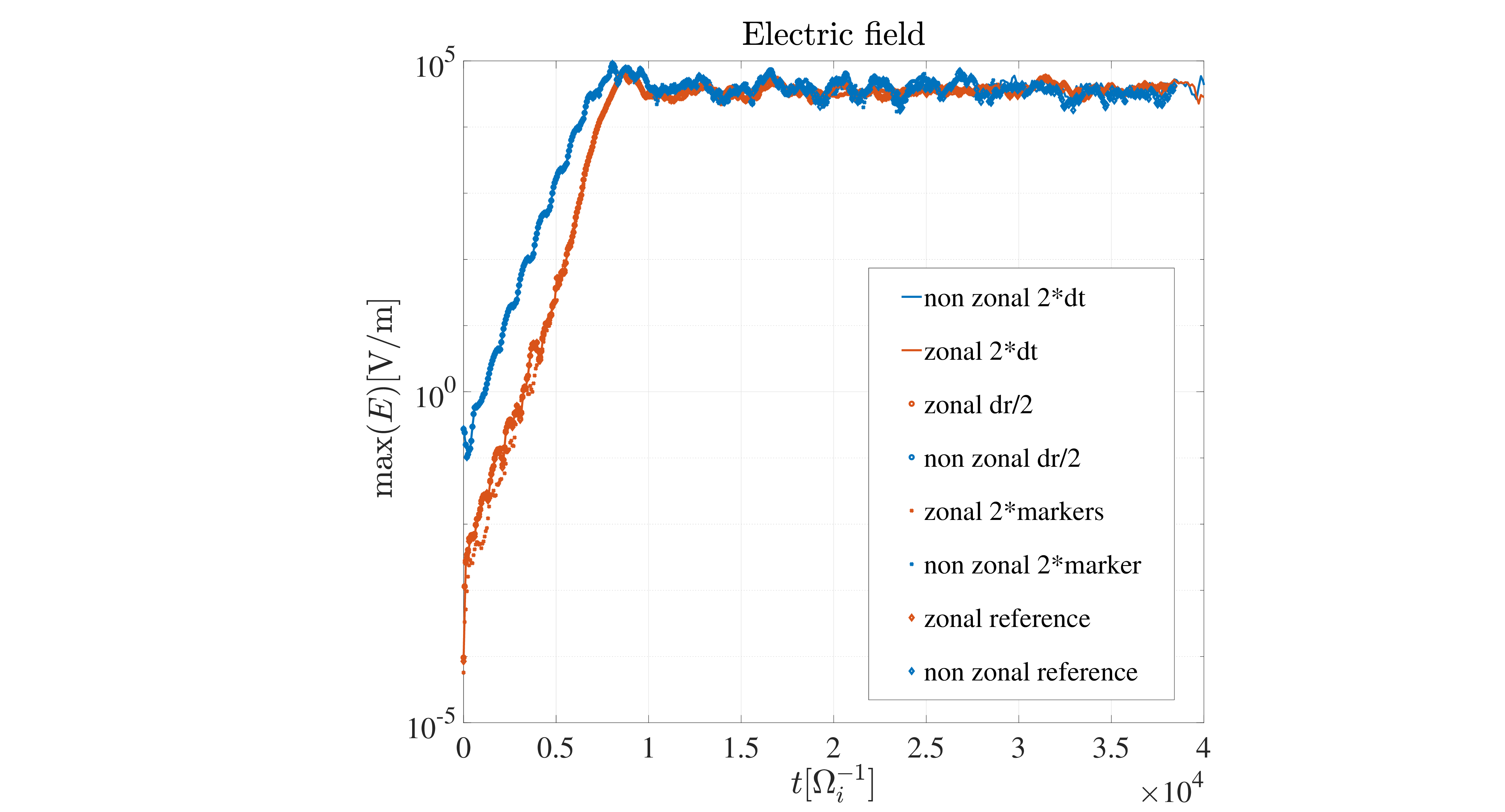}
 \caption{Convergence test }
     \label{appen1}
\end{figure}
We performed a convergence test on the simulation parameters. $dr$ and $dt$ are the step heights of the radial and temporal grid, respectively. A test was also made on the number of markers used for each species. We can conclude from this test that the simulation results are independent of the parameters used for the simulation, as illustrated in \ref{appen1}. 
\end{appendices}

\bibliographystyle{unsrt}
\bibliography{biblio}
\end{document}